\documentclass[twocolumn]{aa}
\usepackage{graphicx}
\usepackage{natbib}
\def\gray {$\gamma$-ray\ }

\def\deg{^\circ}

\def\regionA{the inner region of the Galaxy ($330\deg<l<30\deg, |b|<10\deg$)}
\def\regionB{the central part of the Galaxy ($350\deg<l<10\deg, |b|<10\deg$)}
\def\regionC{the Galactic-disk region ($330\deg<l<350\deg, 10\deg<l<30\deg, |b|<10\deg$)}

\def\regionE{$330\deg<l<30\deg,    |b|< 5\deg$  }
\def\regionRevnivtsev{$350\deg<l<10\deg,  2<|b|<10\deg$  }

\def\Rjk{R_{jk}}
\def\Ij{I_j}

\def\dk{d_k}
\def\bk{b_k}
\def\nk{n_k}

\def\thetai{\theta_i}
\def\thetaj{\theta_j}
\def\thetap{\theta_p}
\def\thetaq{\theta_q}

\def\dthetap{\theta_p-\theta_p^o}
\def\dthetaq{\theta_q-\theta_q^o}

\def\sigmaij{\sigma_{ij}}
\def\sigmapq{\sigma_{pq}}
\def\sigmapp{\sigma_{pp}}

\def\Mij{M_{ij}}
\def\Bik{B_{ik}}
\def\Sik{S_{ik}}

\def\NI{N_I}
\def\NB{N_B}

\def\logL{\log L}
\def\Hessian{H}
\def\Hessianpq{H_{pq}}
\def\Hessianpp{H_{pp}}

\begin{document}
\title{Gamma-ray continuum emission from the inner Galactic region as observed with INTEGRAL/SPI}

   \subtitle{ }

   \author{A. W. Strong     \inst{1},
           R. Diehl         \inst{1},
           H. Halloin       \inst{1},
           V. Sch\"onfelder \inst{1},
\\ 
           L. Bouchet       \inst{2},
           P. Mandrou       \inst{2},
           F. Lebrun        \inst{4}
\and
           R. Terrier       \inst{3,4}
%          ...\inst{x,y},
          }

\offprints{A. W. Strong, aws@mpe.mpg.de}

\institute{ Max-Planck-Institut f\"ur extraterrestrische Physik,
Postfach 1312, D-85741 Garching, Germany           
          %    \email{aws@mpe.mpg.de,rod@mpe.mpg.de}
\and
CESR-CNRS, 9 Av. du Colonel Roche, 31028 Toulouse Cedex 04, France
\and
Astroparticule et Cosmologie, 11 place Marcellin Berthelot, 75005 Paris,
France
\and
DSM/DAPNIA/SAp, CEA-Saclay, 91191 Gif-sur-Yvette, France
             }

\date{Received 8 July 2005 / Accepted 28 July 2005 }

\abstract{
The diffuse continuum emission from the Galactic plane in
the energy range 18-1000 keV has been
studied using 16 Ms of data from the SPI instrument on INTEGRAL.
 With such an exposure we can exploit the  imaging properties of SPI to achieve a good separation of
 point sources from the  various diffuse components.
 Using a candidate-source 
catalogue derived with IBIS on INTEGRAL
and  a number of sky distribution  models we obtained spectra resolved in Galactic longitude.
 We can identify 
spectral components of a diffuse continuum
of power law shape with index about 1.7, a positron annihilation component with a continuum 
from positronium and the line at 511 keV,
and a second, roughly power-law component from  detected point sources.
 Our analysis confirms the concentration of positron
annihilation emission in the inner region ($|l|<10^o$), the disk  ($10^o<|l|<30^o$)  being at 
least a factor 7 weaker in this emission.
The power-law component in contrast drops by only a factor 2, showing a quite
different longitude distribution and spatial origin.
%The diffuse power-law component is hard, with spectral index about -1.8.
Detectable sources constitute about 90\% of the total Galactic emission
between 20 and 60 keV, but have a steeper spectrum than the diffuse emission,
their contribution to the total emission dropping rapidly to a small fraction at higher energies.
The spectrum of diffuse emission is compatible with RXTE and COMPTEL at lower
and higher energies respectively.
In the SPI energy range the flux is lower than found by OSSE,
probably due to the more complete accounting for sources by SPI.
The power-law  emission is difficult to explain as of interstellar origin,
 inverse Compton giving  at most 10\%,    and instead
a population of unresolved point sources is proposed as a possible origin,
 AXPs with their spectra hardening above 100 keV  being plausible candidates.
We present a broadband spectrum of the Galactic emission from
10~keV to 100~GeV.

   }
\titlerunning{Gamma-ray continuum emission from the inner Galaxy}
\authorrunning{Strong A.W. et al.}
\maketitle
%
%________________________________________________________________

\section{Introduction}

Interstellar gamma-ray emission from the Galaxy is a major
study objective for INTEGRAL.
The inner Galactic ridge is known to be an intense source of 
continuum hard X- and soft \gray emission: hard X-ray emission was
discovered in 1972 \citep{bleach72}, and interstellar emission has subsequently been observed
from keV to MeV energies by ASCA, Ginga, RXTE, OSSE, COMPTEL and most
recently by Chandra and XMM-Newton.
Most directly comparable to our INTEGRAL analysis are results from OSSE on the Compton
Gamma Ray Observatory \citep{purcell96,kinzer99,kinzer01}.
%\ \citet{purcell96} used coordinated observations of sources in
% the Galactic centre region from the SIGMA satellite to evaluate the source contribution. They concluded that after correction for these sources,
%the OSSE flux was consistent with that measured by OSSE at
% l=25$\deg$ and 339$\deg$.
%\citet{kinzer99,kinzer01} give an OSSE spectrum for $l=b=0$,  but without independent
% imaging observations, so that the point-source contribution remains uncertain.
 
Continuum emission of diffuse, interstellar nature is expected in the hard-X and \gray regime
from the physical processes of positron annihilation (through intermediate formation of positronium atoms)
and of inverse-Compton emission or bremsstrahlung from cosmic-ray electrons.
For the non-positronium continuum, hard X-rays from bremsstrahlung emission imply a luminosity in 
 cosmic-ray electrons which is unacceptably large \citep[see e.g.][]{dogiel02a}.
Composite models have been proposed, with thermal and nonthermal components
 from electrons accelerated in supernovae or ambient interstellar turbulence, by \citet{valinia00b}.
An alternative solution to the luminosity problem has been proposed by \citet{dogiel02b}.
At MeV energies the origin of the emission is also uncertain \citep{SMR00}.

Alternatively, the origin of the ridge emission could be attributed to 
a population of sources too weak to be detected individually, and hence it could be not truly
interstellar. Gamma-ray telescopes in general have inadequate spatial resolution to clarify this issue.  
For X-rays, important progress in this area was made by ASCA \citep{kaneda97} and Ginga \citep{yamasaki97}.
More recently, high-resolution imaging in X-rays (2--10 keV) with Chandra \citep{ebisawa01,ebisawa05} 
has proven the existence of a truly diffuse component; 
similarly, it has been inferred from XMM-Newton \citep{hands04} that
80\% of the Galactic-ridge X-ray emission is probably diffuse, and only 9\% can be
accounted for by Galactic sources (the rest being extragalactic sources).

The study of the Galactic-ridge continuum emission is a key goal of the INTEGRAL mission.  
The high spectral resolution combined with its imaging capabilities promises new 
insights into the nature of this enigmatic radiation.
Furthermore, reliable modelling of the diffuse emission
will be essential for the study of point sources in the inner Galaxy,
since it constitutes a large and anisotropic background against which sources must be resolved.

Previous work based on first, smaller sets of INTEGRAL/SPI observations have reported
the detection of diffuse emission at a level consistent with previous experiments \citep{strong03a,strong03b}.
But statistical and systematic errors were large, due in part to the uncertainty in the point-source
contribution.  
Meanwhile, a new analysis of INTEGRAL/IBIS data \citep{lebrun04,terrier04} showed
 that, up to 100 keV, indeed a large fraction of the total  emission from the inner
Galaxy is due to sources. 
\citet{strong04} used the source catalogue from this work (containing 91 sources)             
as input to     SPI model fitting, giving a much more solid basis %\citep[see also][]{strong04} 
for the contribution of point sources in such an analysis.
 This exploited the complementarity of the  instruments on
INTEGRAL for the first time in the context of diffuse emission.
In the present work we extend this analysis to  include a much larger observation database, and use the
2nd IBIS Catalogue to provide source candidates.

We also draw attention to a related paper \citep{bouchet05}, which describes an independent but similar study
of SPI data, but focussing more on the point-source contribution.

%----------------------------------------------------------------------
\section{INTEGRAL Observations}

The INTEGRAL Core Program \citep{winkler03} for A01 includes the Galactic Centre Deep Exposure (GCDE)
which maps the inner Galaxy ($330\deg<l<30\deg, -20\deg<b<20\deg$) with
a viewing time of about 4Ms per year. The full region is covered in one GCDE cycle,
and there are two cycles per year. Subsequent AOs have varied this strategy somewhat.
 In addition the Galactic Plane Survey (GPS) covers the whole plane at low Galactic latitude, 
with lower exposure than the GCDE. Data from the first three years, including GCDE, GPS,
 data which has become public up to March 2005, and Open Time data for which permission
 has been obtained from the corresponding Principal Investigator, are  used for the study   reported here. 

\begin{figure}
  \centering    
  \includegraphics[width=10cm]{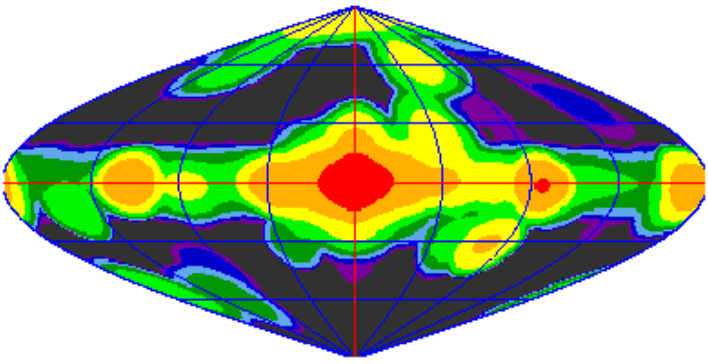}
  \caption{Sky distribution of exposure, at 1 MeV.  Galactic coordinates
centred on $l=b=0$. Units:$10^7$ cm$^2$ s.
Green: 0.5--1.4,  yellow: 1.4--4, orange: 4--13, red: 13--35  }
  \label{exposure}
  \end{figure}

We use data from the SPI (INTEGRAL Spectrometer) instrument;
descriptions of the instrument and performance are given in \citet{vedrenne03,attie03}.  The energy range covered by
SPI is 20~keV -- 8~MeV, but here we restrict the analysis to energies
up to 1 MeV; above these energies the statistics are small and the
analysis is more difficult, so is reserved for future work.  
The data were pre-processed using the INTEGRAL Science Data Centre (ISDC) Standard Analysis   software (OSA) up to
the level containing binned events, with 0.5 keV bins, with pointing and livetime information.
GCDE data from about 200 orbital revolutions from  15 -- 259 were used.  8100 pointings were used; 
the sky exposure  is shown in Fig \ref{exposure}; the total  exposure livetime is $1.6\times 10^7$ s. 
The exposure per pointing is typically 1800 s.
%, but can be as low as 300s in cases of high telemetry losses in the early part of the GCDE.
  The energy calibration is performed using instrumental
background lines with known energies; while this is a critical operation
for line studies (where sub-keV accuracy is required), for continuum
studies a standard calibration ($\sim$1 keV accuracy) is quite adequate.
Various energy binnings were used, depending on the available statistics as a function of energy.
  Only single-detector events are used here, since they dominate below 1 MeV.
 After the failure of detector 2 at IJD\footnote{IJD=INTEGRAL Julian Date starts at 1 Jan 2000}  1434
 the remaining 18 detectors were used,
and after the failure of detector 17 at IJD 1659 the remaining 17 were used.

 The instrumental response is based on extensive Monte Carlo simulations and
parameterization \citep{sturner03}; this has been
tested on the Crab in-flight calibration observations and shown to be
reliable to better than  20\% in absolute flux at the current state of the analysis.

Since INTEGRAL data are dominated by instrumental background, the
analysis needs to have good background treatment methods. In the present
work the background ratios between detectors are  obtained by averaging the entire
dataset over time; this has been found to give results which hardly differ from
taking `OFF' observations, and has the advantage of much higher statistics.
 A check on the systematics in this  background approach was given in \citet{strong03a}.

%----------------------------------------------------------------------
%\section{Data Processing}
%Data synthesis, binning, selection.
%preprocessed data from ISDC - standard processing at MPE. Combining, rebinning in energy.
%Results all FITS files. Available on request. 
%APPENDIX Table of runs and data, parameter summaries. Full spectra tables.
%Table: energy range, sky region, theta per model component, flux/intensity,
% error estimates from Hessian  (per model component and summed)
%%%%%%%%%%%%%%%%%%%%%%%%%%%%%%%%%%%%%%%%%%%%%%%%%%%%%%%%%%%%%%%%%%%%%%%%%%%%%%%%%%

\section{Overview of Methods}

\subsection{Data combination, energy binning, selection} 
The program {\it spiselectscw} combines the observation-by-observation standard-processed data into a single
set of datasets suitable for input to analysis software. Its functions include rebinning in energy,
rejection of anomalous data (e.g. high rates due to solar flares), and detector selection (singles, multiples).
This step is required to reduce the enormous data volume of the  standard-processed data (with its 0.5 keV binning) into
a manageable form.
%%%%%%%%%%%%%%%%%%%%%%%%%%%%%%%%%%%%%%%%%%%%%%%%%%%%%%%
\subsection{Background template generation}
The program {\it spioffback} generates a background template by averaging the detector ratios over all pointings
(either from a set of high-latitude `OFF' observations or from the ON observations themselves),
and using these ratios to compute predicted counts using the livetime information. The total predicted counts
are normalized to the total observed counts in each energy interval: this normalization is for convenience,
but is not essential since the absolute background level is determined later in the fitting procedure.
The template is generated separately for each of the periods separated by detector 2 and 17 failures,
since these failures have a signficant impact on the detector ratios.
%%%%%%%%%%%%%%%%%%%%%%%%%%%%%%%%%%%%%%%%%%%%%%%%%%%%%%%%%%%%%
\subsection{Model fitting}
The program {\it spimodfit} uses the binned count data, pointing and exposure information, 
and the instrumental response, to fit sky models, point sources and background.
The output is in the form of FITS files containing, for all processed energies: the fit parameters,
the covariance matrices and the input models.
% Output from the MCMC chains are also stored.
 {\it spimodfit} is a development (by H. Halloin, MPE)  from  {\it spidiffit}
 which was used in earlier work \citep{strong03a,strong04}. 
 It includes more flexible background fitting, optimized fitting routines and memory  use,
 and improvement formatting of the output. We use  {\it spimodfit} version 2.4 in this work.

The model skymaps to be fitted are generated with the program  {\it gensky},
which includes  HI and CO survey data,  as well as Gaussians of arbitrary position and width.
 The input source catalogue is in the ISDC  catalogue format.
%%%%%%%%%%%%%%%%%%%%%%%%%%%%%%%%%%%%%%%%%%%%%%%%%%%%%%%%%%%%%
\subsection{Spectra}
The appropriate size of the energy bins is a trade-off between
the high instrumental resolution and the available statistics.
Since this cannot be predicted {\it a priori},
we make the fits in a range of binsizes over the whole spectrum,
and choose for presentation those for which a satisfactory signal-to-noise
is achieved. Hence the binsize increases from $\sim$10 keV at low energies
 to $\sim$100 keV at high energies.
%For special regions like the 511 keV line and the positronium edge,
%the binsize is chosen appropriately.

The spectral fitting is done using the full energy redistribution matrix
in the form of `RMF' and `IRF' SPI response files (see Section 2 for references).
The procedure used is rigorous for the case where the ratio between
the three  components of the IRF (photopeak and the two Compton components) 
is independent of incidence direction: this variation is small
 \footnote{The  rms deviation of these component ratios with direction,
  weighted by the sensitive area in that direction, is at the few percent level relative to the total.
  % galplot_linux/plots/SPI_response_stats_borderjk10
  For example at 500 keV: photopeak: $0.675\pm0.025 $, Compton-1: $0.124\pm0.0066 $  , Compton-2:  $0.202\pm0.03 $. }
so the procedure is accurate.
%The energy response has a significant effect on the fit as can be seen
%from Table 3 which compares the fit using a diagonal response with that for
%the full response.
The energy response has a significant effect on the fits as can be seen
from the difference between the raw data points and the fitted spectra 
(Figs~\ref{spectrumA}--\ref{spectrumC}).
 The effect is mainly on the level of the power-law
component; for the positronium the effect is small.

%%%%%%%%%%%%%%%%%%%%%%%%%%%%%%%%%%%%%%%%%%%%%%%%%%%%%%%%%%%%%%%%%%%%%%%%
\subsection{Evaluation and plotting}
The program {\it galplot} combines the results from many   {\it spimodfit}  runs
into spectra integrated over sky regions, and sources.
It evaluates the fluxes and error bars using the fit  parameters covariance matrices.
%It also evaluates the MCMC chains.
It plots the SPI spectra together  with results from other instruments like RXTE, COMPTEL, EGRET,
and theoretical models from  the {\it galprop} project \citet{SMR04a,SMR04b,SMR04d}.

%%%%%%%%%%%%%%%%%%%%%%%%%%%%%%%%%%%%%%%%%%%%%%%%%%%%%%%%%%%%%%%%%%%%%%%
\section{Model Fitting          }
\subsection{Basic approach}
We distinguish image space and data space in the usual way,
and define the instrument response as the relation between them.
The image is $\Ij$ and the expected data is $\dk$. 
The expected background is $\bk$
Let $\Rjk$ be the response of data element $k$ to image element $j$.
Then
$$\dk = \sum_j \Rjk\Ij + \bk$$
For the the special case of an image component consisting of a  point source, only the position is required
and computing $\dk$ requires no convolution.

An {\it image model} is a parameterized algorithm for composing an image from components. 
For a linear model,
$$\Ij = \sum_i\thetai\Mij$$
where $\thetai$ are the model parameters.

 More generally, the image will still be described by a sum of components,
but the image components will be non-linear functions of the parameters
(e.g. Gaussian, exponential) and each component is described by several parameters $\overline\thetai$:
$$\Ij=\sum_i\Mij(\overline\thetai)$$

Similarly the background can be constructed from components of a background model $\Bik$
$$\bk = \sum_i\thetai\Bik$$
where $\thetai$ now introduces background parameters.
The sums in the above expressions are over the appropriate subsets
of parameters for image and background model respectively.
In this way we can treat image and background model in the same
way in the subsequent analysis, and $\thetai$ includes both. 
The only formal difference between image and background model is that
the image is convolved with $\Rjk$ and the background is not:
$$\dk = \sum_j \Rjk \sum_{i=1}^{i=\NI}\thetai\Mij+\sum_{i=\NI+1}^{i=\NI+\NB}\thetai\Bik $$
where there are $\NI$ image components and $\NB$ background components, and  $N=\NI+\NB$.

In our modelling approach, the measured signal is represented through model
components:
$$\dk = \sum_i \thetai\Sik$$
where the sky part (diffuse + sources) is:
$$\Sik=  \sum_j \Rjk \Mij, \ \ \ \ i=1,\NI$$
and the instrumental-background part is:
$$\Sik = \Bik, \ \ \ \ i=\NI+1,\NI+\NB$$

%\footnote{The procedure above is for a single energy and hence a implies a diagonal response in energy space. 
%However it would be straightforward  to generalize to the case of a dataset
%including many energy channels and parameters for each energy, so that the solution 
%constitutes an {\it energy deconvolution}. In this case
%$\Sik$ includes the off-diagonal response terms.}
% From the point of view of
%the analysis technique there is no difference, just $\Sik$ is larger (by a factor 
%equal to the square of the number of energy channels), and  $\dk$ and $\thetai$ 
%are correspondingly expanded.

Details of the statistical model fitting method are given in Appendix A.

%%%%%%%%%%%%%%%%%%%%%%%%%%%%%%%%%%%%%%%%%%%%%%%%%%%%%%%%%%%%%%%%%%%%%%%%%%%%%%%%%%%%%%%%%%%5
%\subsection{MCMC method}
%Instead of making the Hessian approximation, a correct sampling of the
%posterior can be made using Monte Carlo Markov Chain methods.
%This is especially useful for evaluating linear or non-linear  combinations of correlated
%components (e.g. the sum over diffuse model components),
%which  more conventional statistical methods cannot address
%(except for linear combinations in the `covariance approximation').

%Normally the components are given a positivity constraint on physical grounds, but this
%may be relaxed as a check on the results.

%%%%%%%%%%%%%%%%%%%%%%%%%%%%%%%%%%%%%%%%%%%%%%%%%%%%%%%%%%%%%%%%%%
%\section{Imaging}. 
%Maximum entropy, background from fitting.
%spiskymax

%%%%%%%%%%%%%%%%%%%%%%%%%%%%%%%%%%%%%%%%%%%%%%%%%%%%%%%%%%%%%%%%%%%%%%%%%%%%%%%%%%%%%%%
\section{The models}
\subsection{Diffuse Galactic emission}
Model maps to represent the emission from cosmic-ray interactions with the gas
are integrated HI and $^{12}$CO as described in \citet{SMR04a}.
Other components, including positronium, positron line,  bulge and Galactic  plane, are represented by 
 a set of Gaussians of various widths summarized in Table 1.
%\ref{components}.
For the positronium and positron line we include   Gaussians with FHWM = 5$\deg$, 10$\deg$  \citep{jean03,knodlseder03}.
Generally only the sum of the components are presented since the 
separate components do not necessarily have physical significance;
the idea is to include sufficient aspects of Galactic structure to be able to reproduce the total emission
with reasonable confidence.
Summing  subsets of components allows a morphological decomposition of the spectrum
for example into disc and bulge emission.
%Relaxing the positivity constraint leads to a mathematical rather than a physical
%representation of the sky, which is however more flexible.

Note that our model components are not orthogonal, but this is not a problem since
our use of the covariance matrix explicitly handles the parameter correlations
and the computation of the errors on linear combinations of components.
%%%%%%%%%%%%%%%%%%%%%%%%%%%%%%%%%%%%%%%%%%%%%%%%%%%%%%%%%%%%%%%%%%%%%%%%%%%%%%%%%%%
% INTEGRAL/SPI spimodfit results 
\begin{table}
\label{components}      % is used to refer this table in the text
\caption{Diffuse model components used in the fitting}
\centering                          % used for centering table
\begin{tabular}{l l l c  }        % left and centered columns (5 columns)
\hline\hline                 
Component    & description& longitude         & latitude \\    % table heading 
             &            & FWHM (deg)        & FWHM (deg)      \\    % table heading 
\hline                   
  1           & HI       &          &          \\
  2           & CO       &          &          \\
  3           & Gaussian & 40       &    5     \\
  4           & Gaussian & 60       &    5     \\
  5           & Gaussian & 80       &    5     \\
  6           & Gaussian &  5       &    5     \\
  7           & Gaussian & 10       &    5     \\
  8           & Gaussian & 20       &    5     \\
  9           & Gaussian & 10       &   10     \\
\hline                                   
\end{tabular}
\end{table}
%%%%%%%%%%%%%%%%%%%%%%%%%%%%%%%%%%%%%%%%%%%%%%%%%%%%%%%%%%%%%%%%%%%%%%%%%%%%%%%%%%%%%%
%%%%%%%%%%%%%%%%%%%%%%%%%%%%%%%%%%%%%%%%%%%%%%%%%%%%%%%%%%%%%%%%%%%%%%%%%%%%%%%%%%%%%%
\subsection{Sources}
The sources fitted are based on a preliminary version of
the 2nd ISGRI Catalogue \citep{bird05}, containing 209 sources.
\footnote{
An analysis was also made with 
 version 20 of the ISDC reference catalogue (available from http://isdc.unige.ch), 
in which sources detected by IBIS, SPI and JEMX  are flagged;
This catalogue includes published sources and those announced ATEL/IAU telegrams.
There are 203  sources flagged as detected by IBIS in this catalogue. The results
for the diffuse emission were not significantly different
to those using the 2nd ISGRI Catalogue.}
Only the source coordinates are used here,
since the flux in each energy range is determined in the fitting procedure
simultaneously with the other components.

At low energies it is known that most of the catalogue sources are detectable
and make a corresponding contribution to the emission. At high energies on
the other hand, only a few sources have a hard enough spectrum to be
of relevance. Including the whole catalogue in fact leads to a `glow' or `cross-talk' phenomenon,
whereby the diffuse emission is attributed to the sum of many catalogue sources
at a very low level ($< 1\sigma$), and the fit cannot distinguish
 between this source glow  and emission from  diffuse components.
As a result in this case we get large error bars on both components,
 with highly anti-correlated fluxes,
while the total flux (diffuse + sources) is still well constrained.
Hence at higher energies the number of sources included must be reduced using
our knowledge of the individual sources from IBIS, SPI and other missions.
Beyond 200 keV only a few sources are required.
A  systematic treatment of this effect is to include only sources above
a given detection threshold in each energy range. We use a 3$\sigma$ threshold, thereby 
including all sources which could contribute while excluding the 
low-level `glow' sources. Reducing this threshold (to 1$\sigma$, 2$\sigma$) allows an estimate of the
sensitivity of the diffuse spectrum to the chosen level.
 The energy bands below 100 keV are only 10 keV wide
 (to exploit the energy resolution of SPI as far as allowed by statistics), 
but use of these bands for the source selection is not sensitive enough and leads
 to extra fluctuations of the number of sources with energy;
hence we use wider bands of 30 keV for this purpose. This reduces such fluctuations to an acceptable level..
Between 100 and 200 keV the source-selection bands
are widened from 40 keV to 60 keV, above 200 keV no widening is required.

Table 2
%\ref{source_sigma}
 shows the number of sources above a range of $\sigma$-levels, as
a function of energy.

%An effective way to probe the influence of including 'superfluous' sources is to
%introduce fake sources, most simply by shifting the source positions in the catalogue
%from their true positions. In this case it is known that the source fluxes do
%not correspond to real sources, so that the flux from the 'glow' can be estimated. 
%A useful diagnostic is to  compare the total spectrum for diffuse + sources included in the
%fit for diffuse only and sources only.

Galactic sources are generally variable, except for some like the Crab nebula.
At high energies where the sources are anyway unimportant, there is no problem,
but at low energies the variability must be accounted for. 
The present analysis can take this into account by allowing the fluxes to
vary on some timescale, with a corresponding increase in the number of free parameters.
Only for strong sources is the variability expected to have an influence on the
diffuse emission parameters. We investigate this by comparing the results
assuming constant sources and various variability timescales.

%For a few strong highly variable sources the timescale is input explicitly.

Table B.1
%\ref{variability}
 shows the effect of the source variability on the determination of
diffuse emission, for various energies and variability timescales. 
An important conclusion is that the effect is  small, 
 in fact it is within the quoted statistical errors
for the case of a constant flux; so assuming
constant source fluxes has no significant influence on the fitted diffuse intensity.
Even allowing for  variability of a few strong sources (taken from \citet{bouchet05}) on the SPI pointing timescale
only changes the diffuse emission results at the 20\%  level, and that only at the lowest energies.
This is not really surprising since the diffuse emission has a very different
signature in the data from any given source, and while an inaccurate source flux
at a given epoch degrades the fit for that period, it has hardly any effect on
the global fit.
Allowing for variability of all sources on timescales above 50 days  tends to
 reduce the diffuse flux slightly at low energies,
  by 50\% in the most extreme case in the lowest energy range,
otherwise by less than 30\%, and always within the quoted errors.
 Above 40~keV there is no effect.
 %To formally include the effect of variability,
%the final diffuse spectra are derived allowing a time variability; we adopt a  scale of 100 days.
%%%%%%%%%%%%%%%%%%%%%%%%%%%%%%%%%%%%%%%%%%%%%%%%%%%%%%%%%%%%%%%%%%%%%%%%%%%%%%%%%
\begin{table*}%[t]  %  *=two columns p=separate page h=here t=top
%\begin{table}[h] 
\label{source_sigma}      % is used to refer this table in the text
\caption{Number of sources detected above various significance levels, as function of energy, for the entire input catalogue. }
\centering                          % used for centering table
%\begin{tabular}{r r r r r r r r r r r r}        % left and centered columns (5 columns)
\begin{tabular}{r r r r r r r r r r }  
\hline\hline                
%Energy range &  1$\sigma$& 2$\sigma$& 3$\sigma$&4$\sigma$&5$\sigma$ &10$\sigma$&20$\sigma$&50$\sigma$&100$\sigma$&200$\sigma$&ID   \\    % table heading 
% (keV)       &            &     &        &    \\    % table heading 
Energy       &  1$\sigma$& 2$\sigma$& 3$\sigma$&4$\sigma$&5$\sigma$ &10$\sigma$&20$\sigma$&50$\sigma$&100$\sigma$   \\    % table heading 
 (keV)       &            &     &        &    \\    % table heading          
\hline                      
\hline 
%  28.0-  38.0  &137&113&102&83&65&41&20&10&7&4&2160  \\ % number(>sigma)
%  38.0-  48.0  &117&87&74&61&51&26&16&6&4&2&2161  \\ % number(>sigma)
%  48.0-  58.0  &83&56&34&25&18&11&5&2&1&1&2162  \\ % number(>sigma)
%  58.0-  68.0  &66&33&20&15&10&5&3&2&1&0&2163  \\ % number(>sigma)
%  68.0-  78.0  &72&46&32&19&17&9&4&2&2&1&2164  \\ % number(>sigma)
%  78.0-  88.0  &61&36&25&17&16&7&4&2&1&1&2165  \\ % number(>sigma)
%  88.0-  98.0  &42&19&14&10&7&3&2&1&1&0&2166  \\ % number(>sigma)
%  98.0- 178.0  &58&30&21&16&12&8&2&2&1&1&2167  \\ % number(>sigma)
% 178.0- 258.0  &38&13&4&2&2&2&2&1&0&0&2168  \\ % number(>sigma)
% 258.0- 338.0  &38&6&3&2&2&2&1&1&0&0&2184  \\ % number(>sigma)
% 338.0- 418.0  &21&4&3&2&1&1&1&0&0&0&2169  \\ % number(>sigma)
% 418.0- 498.0  &21&1&1&1&1&1&0&0&0&0&2170  \\ % number(>sigma)
% 518.0-1018.0  &17&1&1&1&1&1&1&0&0&0&2185  \\ % number(>sigma)
\hline 
  18.0-  48.0  &145&134&123&114&93&59&33&15&8   \\ % number(>sigma) 2382
  28.0-  58.0  &145&123&97 &82 &71&39&20&9 &6      \\ % number(>sigma) 2383
  38.0-  68.0  &123&84 &62 &47 &40&17&12&4 &2  \\ % number(>sigma) 2384
  48.0-  78.0  &108&63 &44 &34 &26&14&6 &3 &2  \\ % number(>sigma)2385
  58.0-  88.0  &100&52&36&29&22&11&6&2&2  \\ % number(>sigma) 2386
  68.0-  98.0  &84&57&41&27&19&12&7&2&2  \\ % number(>sigma)  2387
  78.0- 108.0  &77&48&30&20&18&9&5&2&2  \\ % number(>sigma)   2388
  88.0- 148.0  &73&38&23&17&14&8&3&2&1  \\ % number(>sigma)   2389
 128.0- 188.0  &48&20&12&8&6&3&2&1&1    \\ % number(>sigma)   2390
 178.0- 258.0  &41&12&4&3&3&2&2&1&0  \\ % number(>sigma)
 258.0- 338.0  &43&9&4&2&2&2&1&1&0  \\ % number(>sigma)
 338.0- 418.0  &31&4&3&2&1&1&1&0&0  \\ % number(>sigma)
 418.0- 498.0  &25&3&1&1&1&1&0&0&0  \\ % number(>sigma)
\hline                                   %inserts single line
\end{tabular}
\end{table*}
%%%%%%%%%%%%%%%%%%%%%%%%%%%%%%%%%%%%%%%%%%%%%%%%%%%%%%%%%%%%%%%%%%%%%%%%%%%%

%%%%%%%%%%%%%%%%%%%%%%%%%%%%%%%%%%%%%%%%%%%%%%%%%%%%%%%%%%%%%%%%%%%%%%%%%%%%
\section{Results}

\subsection{Background}
\begin{figure}
  \centering    
  \includegraphics[width=6cm]{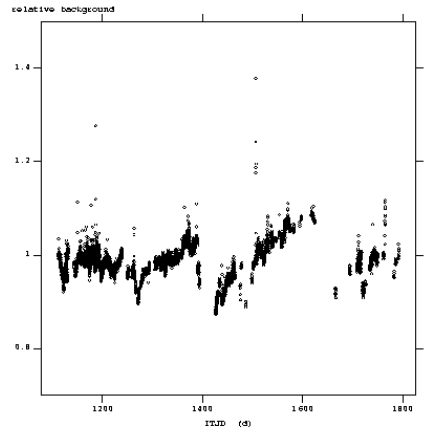}
  \caption{Time dependence of background  for  48 - 58 keV  determined from the fitting procedure.
   Relative units.  Time axis in days since 1 Jan 2000  }
  \label{background}
  \end{figure}

Fig \ref{background} shows the relative background variation determined in the fitting,
per pointing, for a typical energy range; the variations are as expected
from tracers such as saturated Germanium rates and the general upward trend
due to the solar cycle.
Discontinuities appear at the times of detector 2 and 17 failures,
as expected. 
 Occasional high values are encountered which
were not high enough to be removed by the preliminary data filtering,
but these are accounted for in the fitting procedure.

%%%%%%%%%%%%%%%%%%%%%%%%%%%%%%%%%%%%%%%%%%%%%%%%%%%%%%%%%%%%%%%%%%%%%%%%%%%%
%%%%%%%%%%%%%%%%%%%%%%%%%%%%%%%%%%%%%%%%%%%%%%%%%%%%%%%%%%%%%%%%%%%%%%%%%%%%
%%%%%%%%%%%%%%%%%%%%%%%%%%%%%%%%%%%%%%%%%%%%%%%%%%%%%%%%%%%%%%%%%%%%%%%%%%%%
%%%%%%%%%%%%%%%%%%%%%%%%%%%%%%%%%%%%%%%%%%%%%%%%%%%%%%%%%%%%%%%%%%%%%%%%%%%%
%%%%%%%%%%%%%%%%%%%%%%%%%%%%%%%%%%%%%%%%%%%%%%%%%%%%%%%%%%%%%%%%%%%%%%%%%%%%
\subsection{Spectra}
%All fit results are given in tabular form in the Appendix. %\ref{fits_A},\ref{fits_sources},\ref{fits_B},\ref{fits_C} .
%%%%%%%%%%%%%%%%%%%%%%%%%%%%%%%%%%%%%%%%%%%%%%%%%%%%%%%%%%%%%%%%%%%%%%%%%%%%
  \begin{figure}
  \centering    
  \includegraphics[width=8cm]{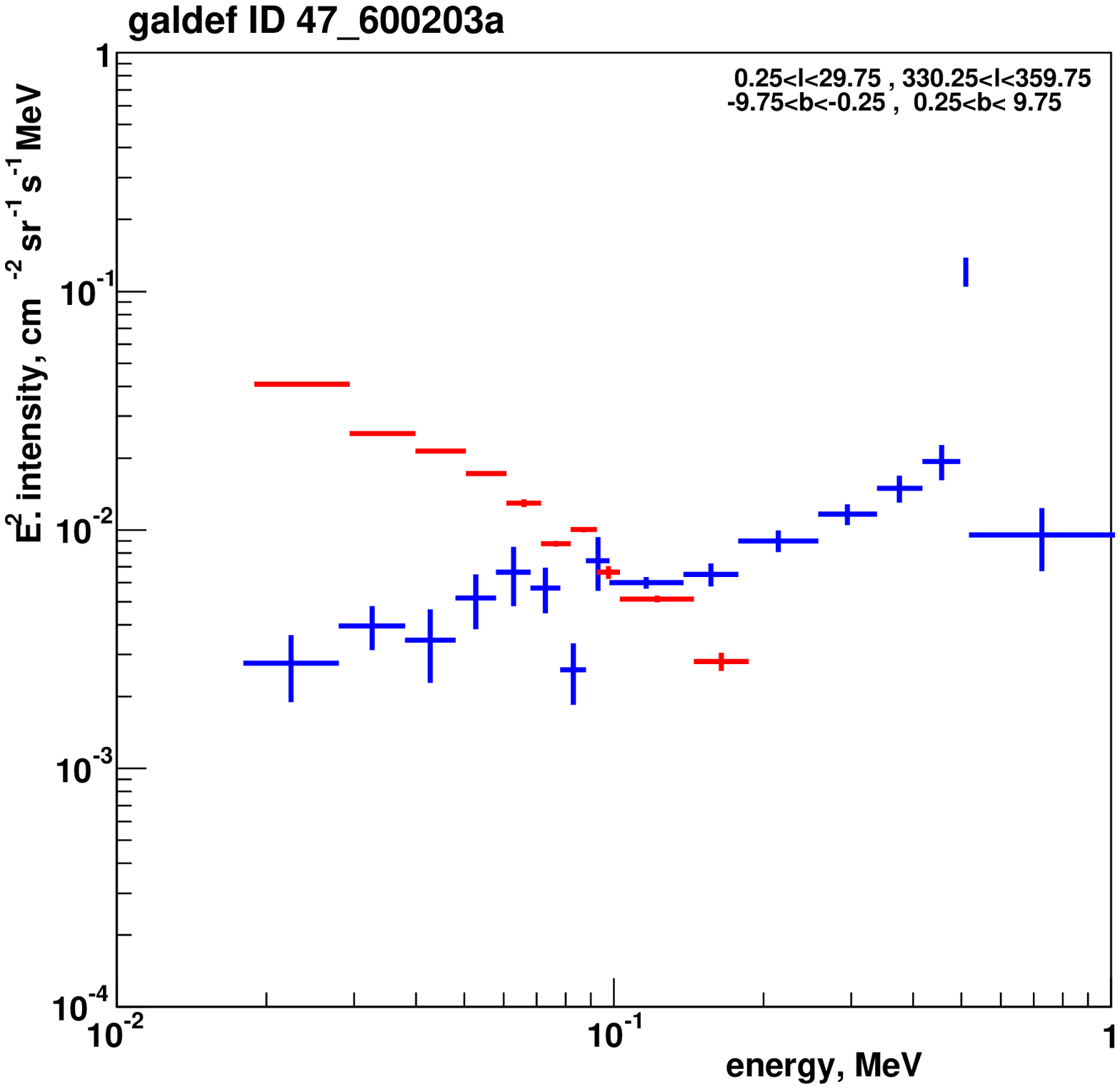}
  \caption{Spectrum of diffuse emission summed over  components (blue), and sum of all sources (red), for \regionA.   
    3$\sigma$   source threshold.    }
  \label{spectrum1}
  \end{figure}
%%%%%%%%%%%%%%%%%%%%%%%%%%%%%%%%%%%%%%%%%%%%%%%%%%%%%%%%%%%%%%%%%%%%%%%%%%%%
%%%%%%%%%%%%%%%%%%%%%%%%%%%%%%%%%%%%%%%%%%%%%%%%%%%%%%%%%%%%%%%%%%%%%%%%%%%%
%  \begin{figure}
%  \centering    
%  \includegraphics[width=8cm]{fig4.eps}
%  \caption{Spectrum of diffuse emission summed over  components (blue), and sum of all sources (red), for \regionA.
%  2$\sigma$   source threshold   }
%  \label{spectrum2sigma}
%  \end{figure}
%______________________________________________________________
%----------------------------------------------------------- 
%  \begin{figure}
%  \centering    
%  \includegraphics[width=8cm]{fig5.eps}
%  \caption{Spectrum of diffuse emission summed over  components (blue), and sum of all sources (red), for \regionA.   
%   1$\sigma$   source threshold     }
%  \label{spectrum1sigma}
%  \end{figure}
%______________________________________________________________

%%%%%%%%%%%%%%%%%%%%%%%%%%%%%%%%%%%%%%%%%%%%%%%%%%%%%%%%%%%%%%%%%%%%%%%%%%%%
  \begin{figure}
  \centering    
  \includegraphics[width=8cm]{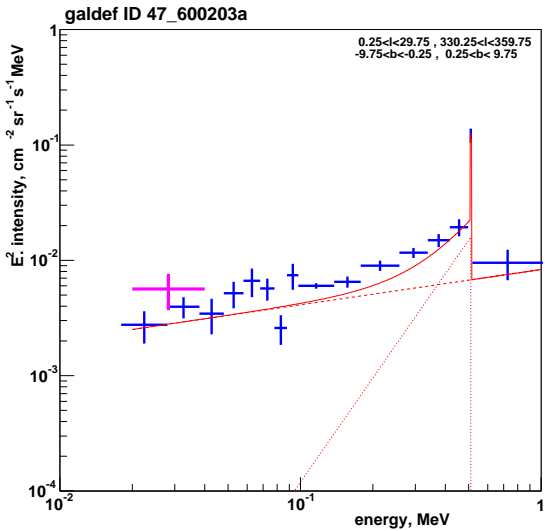} 
  \caption{Spectrum of diffuse emission summed over model components, for \regionA.
  3$\sigma$   source threshold.
Also shown is the fit of SPI data to power-law plus positronium plus line,  using the full energy response.  
Note that the fit lies below the data points because of the non-diagonal energy reponse of SPI.
Magenta point: diffuse emission measured by IBIS/ISGRI  \citep{terrier04}.      }
  \label{spectrumA}
  \end{figure}
%%%%%%%%%%%%%%%%%%%%%%%%%%%%%%%%%%%%%%%%%%%%%%%%%%%%%%%%%%%%%%%%%%%%%%%%%%%%
%%%%%%%%%%%%%%%%%%%%%%%%%%%%%%%%%%%%%%%%%%%%%%%%%%%%%%%%%%%%%%%%%%%%%%%%%%%%
  \begin{figure}
  \centering    
  \includegraphics[width=8cm]{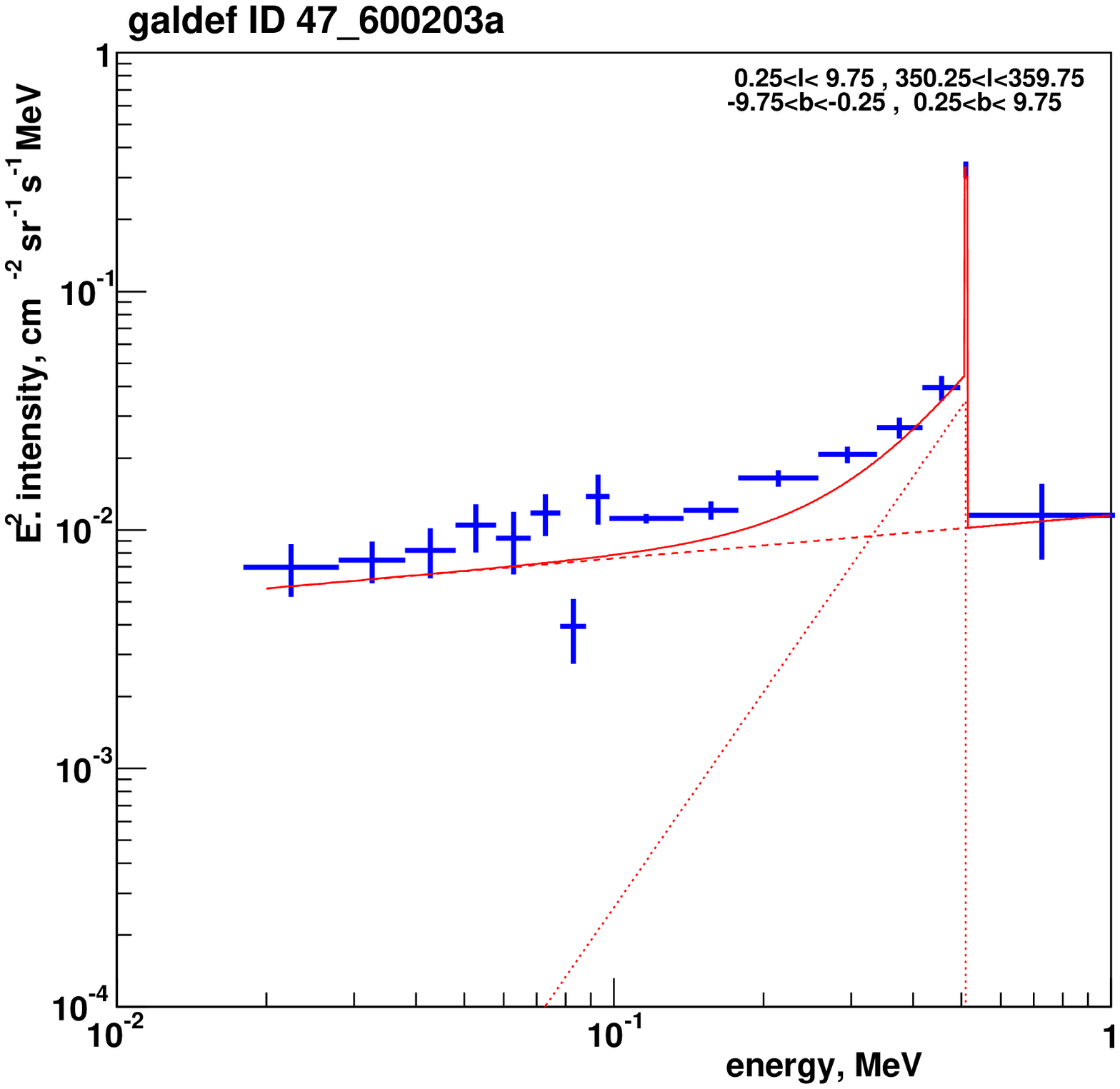}
  \caption{Spectrum of diffuse emission summed over model components, for \regionB. 
  See Fig~4 caption for  details of fit.
        }
  \label{spectrumB}
  \end{figure}
%%%%%%%%%%%%%%%%%%%%%%%%%%%%%%%%%%%%%%%%%%%%%%%%%%%%%%%%%%%%%%%%%%%%%%%%%%%%
%%%%%%%%%%%%%%%%%%%%%%%%%%%%%%%%%%%%%%%%%%%%%%%%%%%%%%%%%%%%%%%%%%%%%%%%%%%%
  \begin{figure}
  \centering    
  \includegraphics[width=8cm]{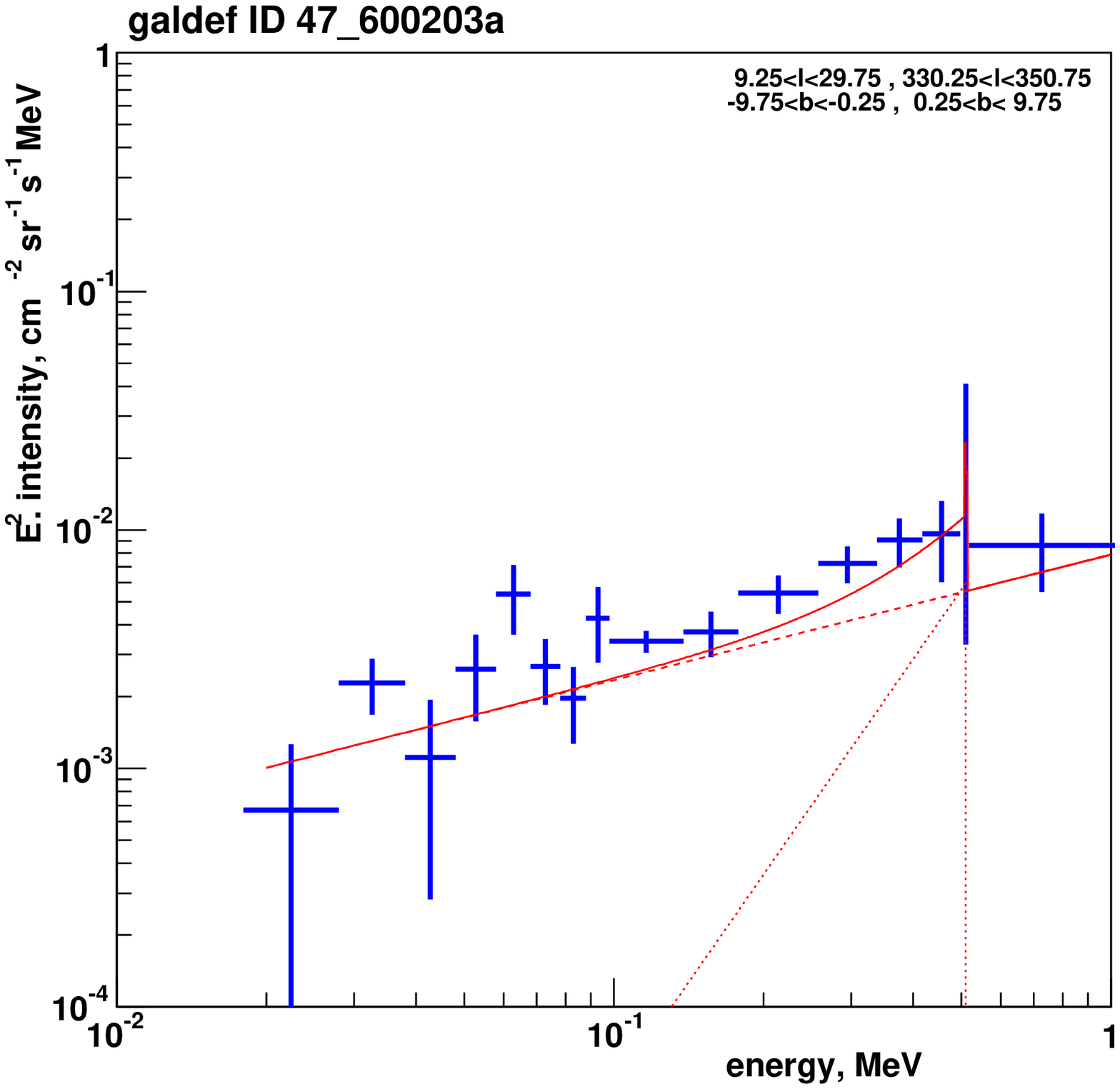}
  \caption{Spectrum of diffuse emission summed over model components, for \regionC.   
  See Fig~4 caption for  details of fit.        }
  \label{spectrumC}
  \end{figure}
%%%%%%%%%%%%%%%%%%%%%%%%%%%%%%%%%%%%%%%%%%%%%%%%%%%%%%%%%%%%%%%%%%%%%%%%%%%%
%%%%%%%%%%%%%%%%%%%%%%%%%%%%%%%%%%%%%%%%%%%%%%%%%%%%%%%%%%%%%%%%%%%%%%%%%%%%
%  \begin{figure}
%  \centering    
%  \includegraphics[width=8cm]{fig7.eps}
%  \caption{Spectrum of diffuse emission summed over model components, for \regionD.          }
%  \label{spectrumD}
%  \end{figure}
%%%%%%%%%%%%%%%%%%%%%%%%%%%%%%%%%%%%%%%%%%%%%%%%%%%%%%%%%%%%%%%%%%%%%%%%%%%%
Fig.~\ref{spectrum1} shows the spectra of the summed catalogue sources and of total diffuse emission,
 for \regionA.
We find a clear change from source-dominated to diffuse-dominated emission in the inner Galaxy,
going from low to high energies.
The transition energy is in the vicinity of 100~keV, but this depends on the fitted sky region
due to variation of relative intensities of the components along the plane of the Galaxy.
The `diffuse' emission may of course include contributions from unresolved sources.
\footnote{In this paper we refer to unresolved emission 
generically as `diffuse', independent of its true nature. }
%Figs \ref{spectrum2sigma} and  \ref{spectrum1sigma} show the effect of reducing
%the threshold for sources included in the fit from 3$\sigma$ to 2$\sigma$ and 1$\sigma$ respectively.

We have studied  the effect of reducing
the threshold for sources included in the fit from 3$\sigma$ to 2$\sigma$ and 1$\sigma$.
As expected the diffuse component decreases while the source component increases.
However the diffuse component decreases by only 30\% for  the 2$\sigma$ source selection,  which
is at the lower limit for a plausible cut. For the 1$\sigma$ source selection, the source
component shows contamination from the diffuse emission above 100 keV due to `cross-talk' between the components.
We adopt the  3$\sigma$ selection as standard in what follows.

% spectral shape results in general
The spectra can be fitted by a combination of power-law and the positron annihilation components (positronium continuum with the
sharp upper edge at 511 keV and a linear decrease towards lower energies, and the 511 keV line).
%(Table \ref{fits_spectrum}).
Fit results for \regionA, \regionB, and \regionC\  are given in Table 3.
Fig.~\ref{spectrumA} shows the spectrum in \regionA, and  Fig.~\ref{spectrumB}\ the spectrum for
 \regionB.
All regions show a hard power law,  with spectral index around 1.7.

%  spectral shape,  the annihilation parts:
The positronium edge at 511 keV is highly significant, and below it the
positronium continuum with its characteristic linear increase (I=AE) is clearly visible.
For \regionC, Fig.~\ref{spectrumC},
the positronium component is small and not significant, as expected, and the spectrum
is consistent with a power law.

Using the fit results for  \regionB,
 the 511 keV flux is             $0.81\pm0.07\times 10^{-3}$ cm$^{-2}$s$^{-1}$, and
 the 3$\gamma$ continuum flux is $4.15\pm0.64\times 10^{-3}$ cm$^{-2}$s$^{-1}$.
This gives a positronium fraction 
\footnote{Positronium fraction defined as $f=2/({3\over2} + {9\over4} r$) where $r$ is the ratio of 
511 keV line flux to 3$\gamma$ continuum flux
\citep{brownleventhal87}} 
$f$ of $0.97-1.08  $, 
consistent with other determinations \citep{kinzer01,jean04,churazov05}; % \citep{Knoedl05,Weiden05,Churazov04}; 
although this is not the main goal of this analysis, it provides a consistency check on the analysis methods.
While the positronium flux decreases by a factor at least $\sim$7 from $|l|<10^o$ to $|l|>10^o$,
the power-law component decreases by only a factor $\sim2$. This clearly shows the difference
in angular distribution and hence a different spatial origin of the emission.

% spectral shape, the powerlaw part
The power-law continuum flux in this inner region is  $1.3 \pm0.25\times 10^{-3}$ cm$^{-2}$s$^{-1}$ for 300--500 keV, i.e. about 1/3 of
the positronium continuum in this range.
The positronium component is small in \regionC, which allows a more 
precise determination of the non-positron continuum.

For comparison the low-energy diffuse emission detection by IBIS/ISGRI \citep{terrier04} is 
also shown in  Fig \ref{spectrumA} ; the agreement is satisfactory.
The SPI spectrum is compared with RXTE, COMPTEL and OSSE results in Fig \ref{spectrum_comparisonB}  ;
the comparison is made in the region \regionB,  since this is most compatible with the
regions reported for  RXTE \citep{revnivtsev03} and OSSE  \citep{kinzer99}.

%%%%%%%%%%%%%%%%%%%%%%%%%%%%%%%%%%%%%%%%%%%%%%%%%%%%%%%%%%%%%%%%%%%%%%%%%%%%%%%%
\begin{table}[h]   % h = here
%\caption{Spectral fit parameters}             % title of Table
\caption{SPI diffuse emission intensity spectrum fitted to \protect\\  
 I(E) = A E$^{-\gamma}$ +  C E (E$<$0.511) + D$\delta$(E-0.511)  cm$^{-2}$ sr$^{-1}$s$^{-1}$ MeV$^{-1}$ 
where E is energy in MeV.
% Without (upper) and with (lower) full spectral response. 
 }
\label{fits_spectrum}      % is used to refer this table in the text
\centering                          % used for centering table
\begin{tabular}{l l l    }        % left and centered columns (3 columns)
\hline\hline                
parameter  & value                   & error     \\    % table heading 
\hline               
 $  330\deg<l<  30\deg$,                        $ |b|< 10\deg$        & &     \\    
\hline 
%A       & 4.14e-03 &  4.5e-04  \\ % SPI spectrum fit
%$\gamma$&     2.17 &     0.03  \\ % SPI spectrum fit
%C       & 2.21e-01 &  2.4e-02  \\ % SPI spectrum fit
%A       & 8.77e-03 &  8.3e-04  \\ % SPI spectrum fit
%$\gamma$&     1.96 &     0.03  \\ % SPI spectrum fit
%C       & 1.30e-01 &  2.6e-02  \\ % SPI spectrum fit

%INTEGRAL/SPI spectral fit for list 212
%A       & 9.51e-03 &  2.1e-03  \\ % SPI spectrum fit
%$\gamma$&     1.82 &     0.10  \\ % SPI spectrum fit
%C       & 1.33e-01 &  3.0e-02  \\ % SPI spectrum fit
% INTEGRAL/SPI spectral fit for list 213

%A       & 1.17e-02 &  1.9e-03  \\ % SPI spectrum fit
%$\gamma$&     1.71 &     0.07  \\ % SPI spectrum fit
%C       & 1.18e-01 &  2.9e-02  \\ % SPI spectrum fit
\hline 
%INTEGRAL/SPI spectral deconvolved fit for list 213
%A       & 7.89e-03 &  1.5e-03  \\ % SPI spectrum fit
%$\gamma$&     1.76 &     0.08  \\ % SPI spectrum fit
%C       & 1.22e-01 &  2.8e-02  \\ % SPI spectrum fit
%INTEGRAL/SPI spectral deconvolved fit for list 214
%A        & 7.85e-03 &  1.5e-03  \\ % SPI spectrum fit
%$\gamma$&     1.77 &     0.08  \\ % SPI spectrum fit
%C        & 1.16e-01 &  2.8e-02  \\ % SPI spectrum fit
%D        & 2.37e-03 &  3.8e-04  \\ % SPI spectrum fit
%INTEGRAL/SPI spectral deconvolved fit for list 215
%A        & 6.91e-03 &  1.3e-03  \\ % SPI spectrum fit
%$\gamma$&     1.84 &     0.08  \\ % SPI spectrum fit
%C        & 1.25e-01 &  2.7e-02  \\ % SPI spectrum fit
%D        & 2.37e-03 &  3.8e-04  \\ % SPI spectrum fit
%INTEGRAL/SPI spectral deconvolved fit for list 215a
%A        & 6.93e-03 &  1.3e-03  \\ % SPI spectrum fit
%$\gamma$&     1.84 &     0.07  \\ % SPI spectrum fit
%C        & 1.22e-01 &  2.6e-02  \\ % SPI spectrum fit
%D        & 2.37e-03 &  3.8e-04  \\ % SPI spectrum fit
% INTEGRAL/SPI spectral deconvolved fit for list 216
%A        & 7.31e-03 &  1.7e-03  \\ % SPI spectrum fit
%$\gamma$&     1.77 &     0.10  \\ % SPI spectrum fit
%C        & 1.27e-01 &  2.9e-02  \\ % SPI spectrum fit
%D        & 2.36e-03 &  3.8e-04  \\ % SPI spectrum fit
%INTEGRAL/SPI spectral deconvolved fit for list 224
%A        & 8.32e-03 &  1.8e-03  \\ % SPI spectrum fit
%$\gamma$&     1.69 &     0.09  \\ % SPI spectrum fit
%C        & 1.19e-01 &  2.8e-02  \\ % SPI spectrum fit
%D        & 2.36e-03 &  3.8e-04  \\ % SPI spectrum fit

A        & 8.32 10$^{-3}$ &  1.8 10$^{-3}$  \\ % SPI spectrum fit
$\gamma$&     1.69 &     0.09  \\ % SPI spectrum fit
C        & 1.19  10$^{-1}$ &  2.8  10$^{-2}$ \\ % SPI spectrum fit
D        & 2.36  10$^{-3}$ &  3.8  10$^{-4}$ \\ % SPI spectrum fit
\hline \hline
$  350\deg<l<  10\deg$,                        $ |b|< 10\deg$        & &     \\    
\hline 
%A       & 2.77e-03 &  4.0e-04  \\ % SPI spectrum fit
%$\gamma$&     2.24 &     0.05  \\ % SPI spectrum fit
%C       & 1.20e-01 &  2.5e-02  \\ % SPI spectrum fit
%A       & 5.48e-03 &  7.4e-04  \\ % SPI spectrum fit
%$\gamma$&     2.05 &     0.04  \\ % SPI spectrum fit
%C       & 6.06e-02 &  2.7e-02  \\ % SPI spectrum fit

% INTEGRAL/SPI spectral fit for list 212
%A       & 1.47e-02 &  2.5e-03  \\ % SPI spectrum fit
%$\gamma$&     1.90 &     0.08  \\ % SPI spectrum fit
%C       & 2.70e-01 &  4.2e-02  \\ % SPI spectrum fit

% INTEGRAL/SPI spectral fit for list 213
%A       & 1.83e-02 &  2.5e-03  \\ % SPI spectrum fit
%$\gamma$&     1.79 &     0.06  \\ % SPI spectrum fit
%C       & 2.43e-01 &  4.1e-02  \\ % SPI spectrum fit
\hline 
% INTEGRAL/SPI spectral deconvolved fit for list 213
%A       & 1.24e-02 &  1.9e-03  \\ % SPI spectrum fit
%$\gamma$&     1.84 &     0.07  \\ % SPI spectrum fit
%C       & 2.44e-01 &  3.8e-02  \\ % SPI spectrum fit
%INTEGRAL/SPI spectral deconvolved fit for list 214
%A        & 1.23e-02 &  1.9e-03  \\ % SPI spectrum fit
%$\gamma$&     1.85 &     0.07  \\ % SPI spectrum fit
%C        & 2.29e-01 &  3.7e-02  \\ % SPI spectrum fit
%D        & 6.65e-03 &  5.6e-04  \\ % SPI spectrum fit
%INTEGRAL/SPI spectral deconvolved fit for list 215
%A        & 8.80e-03 &  1.4e-03  \\ % SPI spectrum fit
%$\gamma$&     2.00 &     0.07  \\ % SPI spectrum fit
%C        & 2.77e-01 &  3.7e-02  \\ % SPI spectrum fit
%D        & 6.64e-03 &  5.6e-04  \\ % SPI spectrum fit
%INTEGRAL/SPI spectral deconvolved fit for list 215a
%A        & 9.31e-03 &  1.4e-03  \\ % SPI spectrum fit
%$\gamma$&     1.99 &     0.06  \\ % SPI spectrum fit
%C        & 2.65e-01 &  3.6e-02  \\ % SPI spectrum fit
%D        & 6.65e-03 &  5.6e-04  \\ % SPI spectrum fit
% INTEGRAL/SPI spectral deconvolved fit for list 216
%A        & 9.56e-03 &  2.0e-03  \\ % SPI spectrum fit
%$\gamma$&     1.92 &     0.09  \\ % SPI spectrum fit
%C        & 2.80e-01 &  4.0e-02  \\ % SPI spectrum fit
%D        & 6.63e-03 &  5.6e-04  \\ % SPI spectrum fit
%INTEGRAL/SPI spectral deconvolved fit for list 224
%A        & 1.15e-02 &  2.2e-03  \\ % SPI spectrum fit
%$\gamma$&     1.82 &     0.08  \\ % SPI spectrum fit
%C        & 2.61e-01 &  4.0e-02  \\ % SPI spectrum fit
%D        & 6.62e-03 &  5.6e-04  \\ % SPI spectrum fit

A        & 1.15 10$^{-2}$&  2.2 10$^{-3}$  \\ % SPI spectrum fit
$\gamma$&     1.82 &     0.08  \\ % SPI spectrum fit
C        & 2.61 10$^{-1}$ &  4.0 10$^{-2}$  \\ % SPI spectrum fit
D        & 6.62 10$^{-3}$ &  5.6 10$^{-4}$  \\ % SPI spectrum fit

\hline  \hline
$  330\deg<l< 350\deg$, $   10\deg<l<  30\deg$            &    &    \\  
$ |b|< 10\deg$                                            &    &    \\  
\hline 
%A       & 8.78e-03 &  9.9e-04  \\ % SPI spectrum fit
%$\gamma$&     2.02 &     0.04  \\ % SPI spectrum fit
%C       & 3.95e-01 &  3.5e-02  \\ % SPI spectrum fit
%A       & 1.66e-02 &  1.5e-03  \\ % SPI spectrum fit
%$\gamma$&     1.85 &     0.03  \\ % SPI spectrum fit
%C       & 2.62e-01 &  3.8e-02  \\ % SPI spectrum fit

%INTEGRAL/SPI spectral fit for list 212
%A       & 7.12e-03 &  2.6e-03  \\ % SPI spectrum fit
%$\gamma$&     1.70 &     0.16  \\ % SPI spectrum fit
%C       & 6.65e-02 &  3.5e-02  \\ % SPI spectrum fit

% INTEGRAL/SPI spectral fit for list 213
%A       & 8.51e-03 &  2.5e-03  \\ % SPI spectrum fit
%$\gamma$&     1.60 &     0.12  \\ % SPI spectrum fit
%C       & 5.90e-02 &  3.3e-02  \\ % SPI spectrum fit
\hline 
%INTEGRAL/SPI spectral deconvolved fit for list 213
%A       & 5.60e-03 &  2.0e-03  \\ % SPI spectrum fit
%$\gamma$&     1.66 &     0.14  \\ % SPI spectrum fit
%C       & 6.54e-02 &  3.2e-02  \\ % SPI spectrum fit

% INTEGRAL/SPI spectral deconvolved fit for list 214
%A        & 5.63e-03 &  2.0e-03  \\ % SPI spectrum fit
%$\gamma$&     1.67 &     0.14  \\ % SPI spectrum fit
%C        & 6.36e-02 &  3.1e-02  \\ % SPI spectrum fit
%D        & 2.75e-04 &  4.3e-04  \\ % SPI spectrum fit
%INTEGRAL/SPI spectral deconvolved fit for list 215
%A        & 7.55e-03 &  2.1e-03  \\ % SPI spectrum fit
%$\gamma$&     1.55 &     0.11  \\ % SPI spectrum fit
%C        & 4.20e-02 &  3.1e-02  \\ % SPI spectrum fit
%D        & 2.76e-04 &  4.3e-04  \\ % SPI spectrum fit
%INTEGRAL/SPI spectral deconvolved fit for list 215a
%A        & 7.20e-03 &  2.0e-03  \\ % SPI spectrum fit
%$\gamma$&     1.56 &     0.11  \\ % SPI spectrum fit
%C        & 4.38e-02 &  3.1e-02  \\ % SPI spectrum fit
%D        & 2.79e-04 &  4.3e-04  \\ % SPI spectrum fit
%INTEGRAL/SPI spectral deconvolved fit for list 216
%A        & 8.49e-03 &  2.5e-03  \\ % SPI spectrum fit
%$\gamma$&     1.44 &     0.13  \\ % SPI spectrum fit
%C        & 4.11e-02 &  3.1e-02  \\ % SPI spectrum fit
%D        & 2.68e-04 &  4.1e-04  \\ % SPI spectrum fit
%INTEGRAL/SPI spectral deconvolved fit for list 224
%A        & 7.87e-03 &  2.5e-03  \\ % SPI spectrum fit
%$\gamma$&     1.47 &     0.14  \\ % SPI spectrum fit
%C        & 4.49e-02 &  3.2e-02  \\ % SPI spectrum fit
%D        & 2.70e-04 &  4.3e-04  \\ % SPI spectrum fit

A        & 7.87  10$^{-3}$ &  2.5 10$^{-3}$  \\ % SPI spectrum fit
$\gamma$&     1.47 &     0.14  \\ % SPI spectrum fit
C        & 4.49  10$^{-2}$ &  3.2  10$^{-2}$  \\ % SPI spectrum fit
D        & 2.70  10$^{-4}$ &  4.3  10$^{-4}$ \\ % SPI spectrum fit

\hline  \hline
%$  300\deg<l< 330\deg$, $   30\deg<l<  60\deg$            &    &    \\  
%$ |b|< 10\deg$                                            &    &    \\  
%\hline
%INTEGRAL/SPI spectral fit for list 213
%A       & 6.50e-03 &  3.7e-03  \\ % SPI spectrum fit
%$\gamma$&     1.37 &     0.22  \\ % SPI spectrum fit
%C       & 7.24e-03 &  4.1e-02  \\ % SPI spectrum fit
\hline 
%INTEGRAL/SPI spectral deconvolved fit for list 213
%A       & 4.59e-03 &  3.0e-03  \\ % SPI spectrum fit
%$\gamma$&     1.40 &     0.24  \\ % SPI spectrum fit
%C       & 1.01e-02 &  4.0e-02  \\ % SPI spectrum fit
% INTEGRAL/SPI spectral deconvolved fit for list 214
%A        & 4.63e-03 &  3.0e-03  \\ % SPI spectrum fit
%$\gamma$&     1.40 &     0.24  \\ % SPI spectrum fit
%C        & 9.59e-03 &  4.0e-02  \\ % SPI spectrum fit
%D        & -7.43e-05 &  5.4e-04  \\ % SPI spectrum fit
% INTEGRAL/SPI spectral deconvolved fit for list 215
%A        & 8.46e-03 &  3.6e-03  \\ % SPI spectrum fit
%$\gamma$&     1.04 &     0.18  \\ % SPI spectrum fit
%C        & -1.05e-02 &  3.7e-02  \\ % SPI spectrum fit
%D        & -8.77e-05 &  5.4e-04  \\ % SPI spectrum fit
%INTEGRAL/SPI spectral deconvolved fit for list 215a
%A        & 7.68e-03 &  2.0e-03  \\ % SPI spectrum fit
%$\gamma$&     1.08 &     0.10  \\ % SPI spectrum fit
%C        & 0.0      &  4.4e-02  \\ % SPI spectrum fit
%D        & 0.0      &  9.7e-04  \\ % SPI spectrum fit

\hline  
\end{tabular}
\end{table}
%%%%%%%%%%%%%%%%%%%%%%%%%%%%%%%%%%%%%%%%%%%%%%%%%%%%%%%%%%%%%%%%%%%

%%%%%%%%%%%%%%%%%%%%%%%%%%%%%%%%%%%%%%%%%%%%%%%%%%%%%%%%%%%%%%%%%%%%%%%%%%%%%%%
%%%%%%%%%%%%%%%%%%%%%%%%%%%%%%%%%%%%%%%%%%%%%%%%%%%%%%%%%%%%%%%%%%%%%%%%%%%%%%%
\section{Discussion}
%%%%%%%%%%%%%%%%%%%%%%%%%%%%%%%%%%%%%%%%%%%%%%%%%%%%%%%%%%%%%%%%%%%%%%%%%%%%%%%%%%%%%
\subsection{Spectral shape}
\begin{figure}[h]
  \centering    
  \includegraphics[width=8cm]{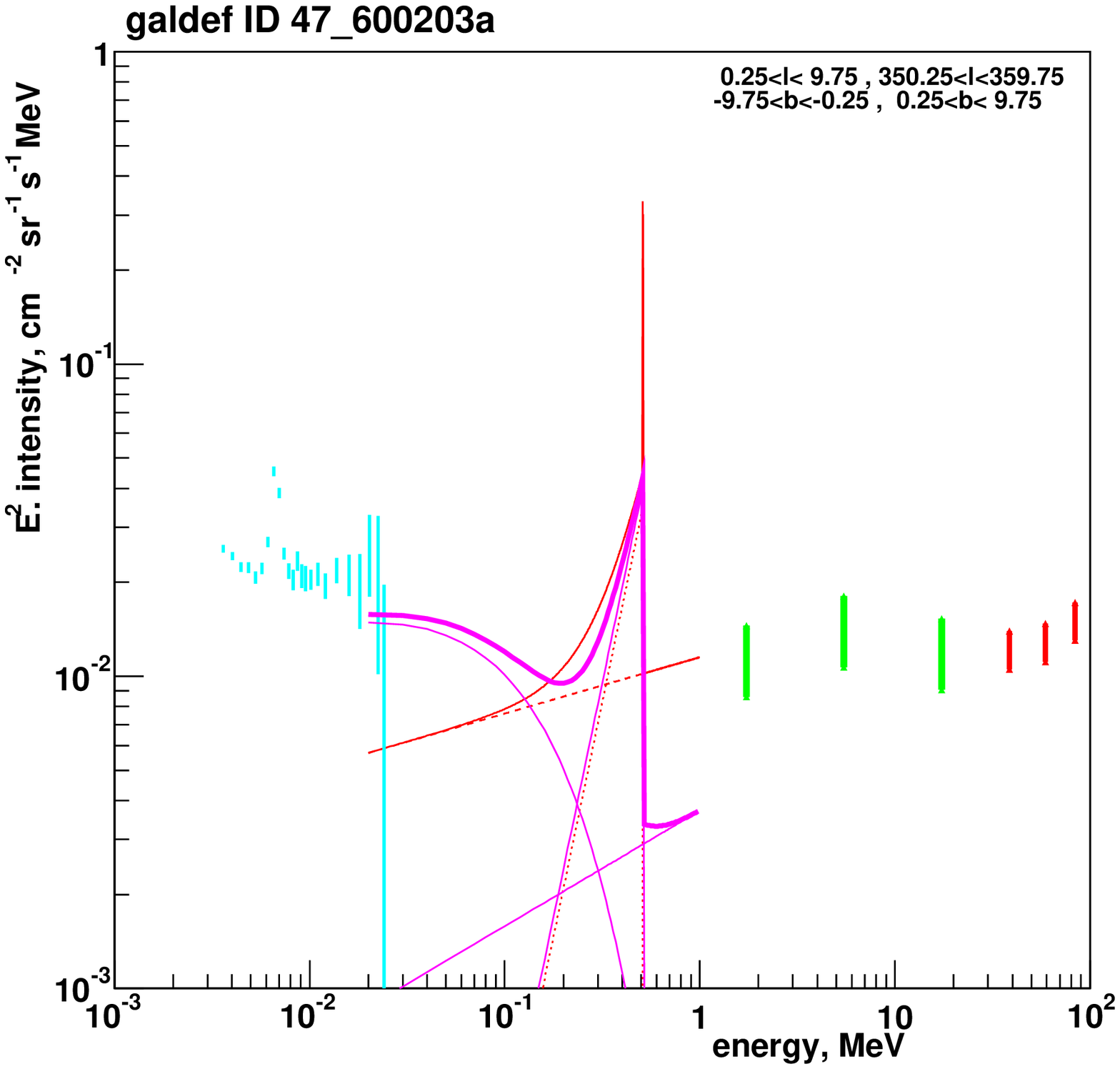}
  \caption{
  Comparison of spectrum of diffuse emission (\regionB),
  with other experiments.
  Data from RXTE (cyan) \citep{revnivtsev03}, 
 SPI (blue) (this work),
% IBIS (magenta cross) \citep{terrier04},
 OSSE (magenta  curves) \citep{kinzer99} (their Table 3 and Fig 5, VP 5+16: $l=0 , b=0$)  ,
  COMPTEL (green)  \citep{strong99}
 and EGRET (red) \citep{SMR04b}.
 Note that the RXTE data  refer to the region \regionRevnivtsev (see text).
    }
  \label{spectrum_comparisonB}
  \end{figure}

% first: comparison to other measurements, spectral shape
The OSSE spectrum is somewhat higher than the SPI result in the 20--100 keV range  (Fig \ref{spectrum_comparisonB}).
This suggests that the OSSE result includes significant source contamination, which is better
accounted for in the SPI analysis.

The RXTE spectrum is certainly above the extrapolation of the SPI power-law.
This suggests a much steeper spectrum below 20 keV, or an additional component with a cutoff around this energy.
For RXTE, we use the results of  \citet{revnivtsev03}
\footnote{The total energy flux in this region in the 3 - 20 keV band is $3.5\times 10^{-9}$ erg cm$^{-2}$ s$^{-1}$
(M. Revnivtsev, private communication) and this is used for normalization of the RXTE spectrum of  \citet{revnivtsev03}.}
 since this is probably  more
free of point-source contamination than e.g. \citet{valinia98} and \citet{valinia00a};
 however it refers to the region \regionRevnivtsev, i.e. avoiding the Galactic ridge itself.
The latitude distribution of diffuse emission  measured by RXTE is rather broad
\citep{valinia98} so the use of this spectrum is appropriate

A rapid softening of the spectrum below the SPI range is  indicated by
a comparison with XMM and Chandra results.
\citet{hands04} (XMM) find a total 2--10 keV intensity $9.6\times 10^{-11}$ erg cm$^{-2}$ s$^{-1}$ deg$^{-2}$ for
$l=19^o - 22^o, |b|\le0.5^o$, 80\% of which is diffuse or at least not in known sources,
corresponding to 0.2~MeV  cm$^{-2}$ s$^{-1}$ sr$^{-1}$ in the units of  Fig \ref{spectrum_comparisonB}.
\citet{ebisawa01,ebisawa05} (Chandra) finds a similar value for diffuse emission at  $l=28^o, b=-0.2^o$.
Even allowing for the large concentration of the emission to the ridge,
 this is above the extrapolation of the SPI power-law by at least a factor 3.
This is consistent with the inclusion of a cutoff power-law at low energies by \citet{kinzer99} in the OSSE analysis.
It is also compatible with  the steeper power-law (index~$>2$)  below 20 keV measured by Ginga \citep{yamasaki97}.

%----------------------------------------------------------- 
%______________________________________________________________

%----------------------------------------------------------- 
  \begin{figure*}%[t] % t=top of page, if not used the tables disappeared at one point!
  \centering    
  \includegraphics[width=10cm]{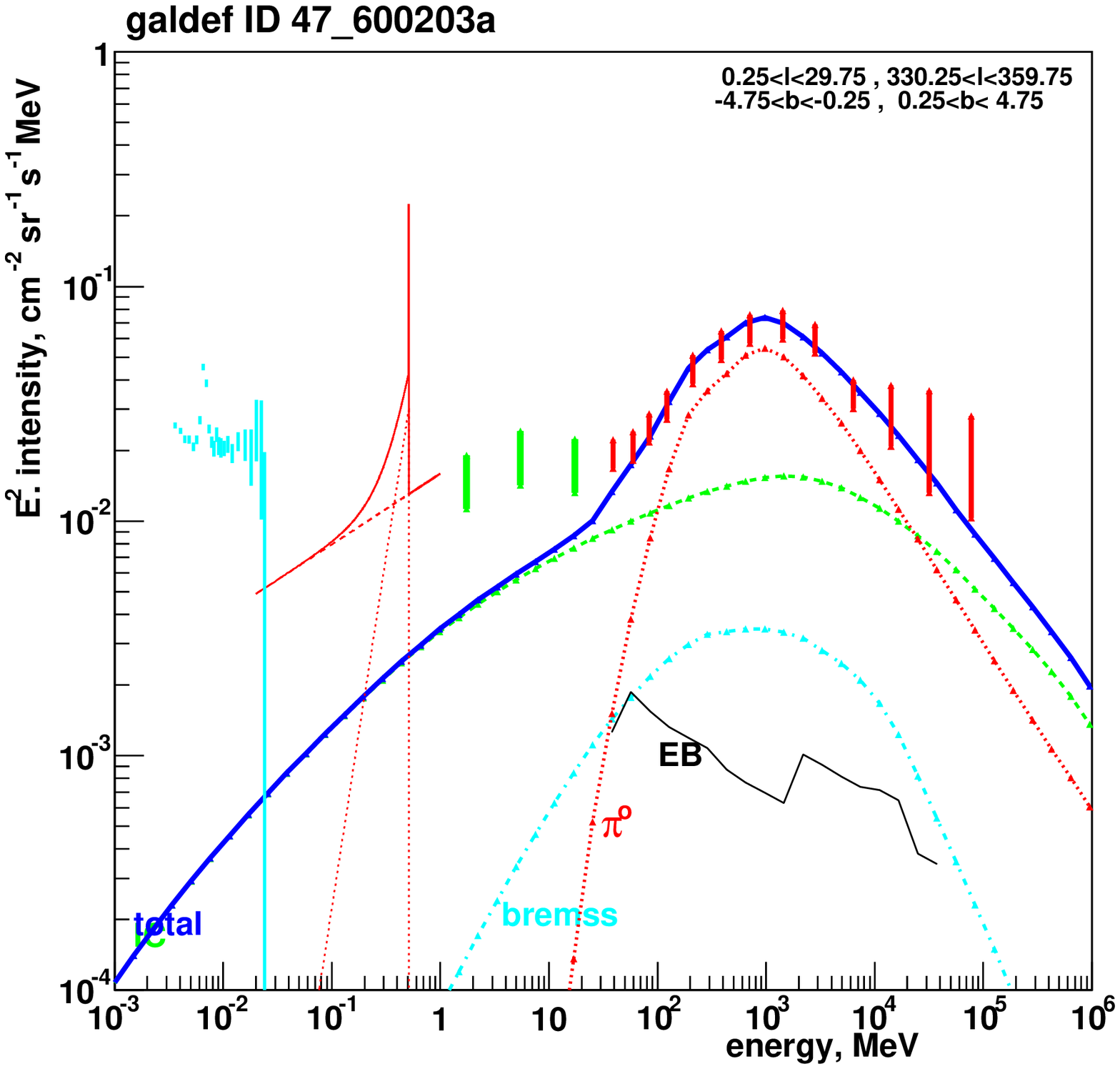}
  \caption{Broadband spectrum of diffuse emision (\regionE),
 compared with model based on cosmic-ray interactions from  \citet{SMR04b,SMR04d}.
  Data from RXTE (cyan),  SPI (red lines), IBIS (magenta), COMPTEL (green),
(references as in  Fig \ref{spectrum_comparisonB})
  and EGRET (red bars) \citep{SMR04b} are shown.  
Dashed/green  curve: inverse Compton,dotted/red: $\pi^o$-decay, dashed-dot/light blue: bremsstrahlung,  
 EB: extragalactic background, dark blue: total model.
Notes: (a) the RXTE data  refer to the region \regionRevnivtsev, so this may explain in part the high intensity
relative to the SPI results;
(b) the EGRET data $>$10 GeV have been corrected for the new sensitivity estimate given by \citet{thompson05},
 which results in higher values than in  \citet{SMR04b}; a 50\% flux error has been adopted to plot this data.
  }
  \label{spectrum_broadband}
  \end{figure*}
%%%%%%%%%%%%%%%%%%%%%%%%%%%%%%%%%%%%%%%%%%%%%%%%%%%%%%%%%%%%%%%%%%%%%%%%%%%%%%%%%%%%%%%

% now: the source/diffuse ratio

%%%%%%%%%%%%%%%%%%%%%%%%%%%%%%%%%%%%%%%%%%%%%%%%%%%%%%%%%%%%%%%%%%%%%%%%%%%%%%%%%%%%%
\subsection{Sources: resolved and unresolved}
For 20-100 keV the resolved source/diffuse emission ratio is about 10  (Fig \ref{spectrum1}), which is compatible
with existing estimates of the total source emission in the Galaxy: \citep{grimm02}
of 2 10$^{39}$ erg s$^{-1}$ and the total diffuse emission of  10$^{38}$ erg s$^{-1}$
\citep{dogiel02a}. These estimates are mainly based on the energy range 2--10 keV (largely RXTE-based)
so this is only a plausibility check.
\footnote{It is important to note the reason for the difference between the `diffuse-dominated'
 Galactic emission  found  by XMM and Chandra 
(Hands et al. 2004, Ebisawa et al. 2005)  and the
`source dominated' emission shown in Fig 3. The latter refers to
a large region of the Galaxy including many strong sources,
while XMM and Chandra refer to a  small field-of-view chosen to
avoid strong sources. Since the emission from the whole Galaxy is
dominated by the brightest sources, the relative importance
of source and diffuse components depends  sensitively on the size of region considered.}

It is of interest to estimate the limitations of the present analysis
which arise from the SPI instrument itself:
the number of sources visible is ultimately limited, at low energies, by source confusion.
 The SPI angular response has  a  width of about 2.6$^o$, which means that
for the inner radian  in longitude within 5$^o$ of the plane we cannot
detect more than $60\times10 / 2.6^2 \sim$100 sources. This limits the flux which could ever
be explicitly attributed to detected sources, as follows.
From Table 2, using the data between 3$\sigma$ and 10$\sigma$
and the first 6 energy ranges,
we find an average logN($>$S)-logS slope of -0.9, i.e. converging to low fluxes.
There are 57 sources  $>3\sigma$ in 18~--~48 keV in the  area chosen above,
and from this it follows that we formally have to go down to 1.6$\sigma$
to reach  the confusion limit.
Integrating N(S)$\times$S, we find that the total source flux increases  by
about 1.5 going from our adopted 3$\sigma$ (Fig 3) to the confusion limit.
This is a very rough estimate, but illustrates that we are near the limit
of what could be deduced from the SPI data, and that the  total flux in {\it detected} sources
 could never be more than 50\% larger than shown in Fig 3.

Sources undetectable because of confusion or low flux will by definition be
interpreted as part of the diffuse emission in our analysis.
The full analysis must then involve detailed modelling of source populations,
which will be addressed in future work.

% now: the diffuse-continuum theory
\subsection{Diffuse continuum emission}
The broad-band spectrum from 2 keV to 100 GeV, including RXTE, SPI, COMPTEL and EGRET, 
for \regionE \footnote{region chosen for compatibility with \citet{SMR04b}}
is compared with  predictions using the {\it galprop} model in Fig \ref{spectrum_broadband}.
%The data from the different experiments are consistent where energy ranges overlap.
The RXTE spectrum is for \regionRevnivtsev so that its inclusion is only indicative,
but illustrates the softening of the spectrum below 20 keV.
The model is the `optimized' one from \citet{SMR04b,SMR04d},
 including the cosmic-ray generated bremsstrahlung and inverse Compton components.
Evidently neither of these processes is sufficient to account for
the observed non-thermal hard X-ray  spectrum; increasing the inverse Compton by invoking
a steeper electron spectrum leads to an overprediction of the EGRET data points.
Alternative explanations of the hard X-ray spectrum have been mentioned in the Introduction,
but these do not easily produce the observed very hard spectrum, so that
a population of compact sources seems to us most likely, and anomalous X-ray pulsars (AXPs), with
their very hard spectra \citep{kuiper04},  appear good candidates.
AXPs also have an appropriate young population distribution compatible with the SPI result.

This picture is consistent with a change of origin from truly diffuse emission below 20~keV
with a steep spectrum, as seen by RXTE, XMM and Chandra,
 to a source-dominated origin with a hard spectrum above 20 keV as seen by SPI.

The present work will form the basis for more detailed evaluation of these models.

\section{Conclusions}
The diffuse continuum emission from the Galactic plane has been detected at high significance with SPI.
The positronium and non-thermal components are clearly separated, with the 
positronium dominating above 300 keV in the inner  Galaxy.
The contribution from detectable point-sources has for the first time been
explicitly accounted for, which was not possible with earlier experiments.
The combined emission from detectable sources varies from a factor 10 times 
the diffuse emission at low energies to negligible at high energies.

Spectral analysis,
using the SPI energy response,  reveals a power-law component
with index about -1.7, consistent with  OSSE and COMPTEL results.
The positronium flux drops by factor at least 7 from $|l|<10^o$ to $|l|>10^o$
in accord with other studies.
The power-law component by contrast drops by only a factor 2, proving a quite
different longitude distribution and spatial origin.

We suggest  a change of origin from truly diffuse emission below 20 keV
with a steep spectrum, 
 to a source-dominated origin with a hard spectrum above 20 keV.

From the broadband spectrum of the diffuse emision from keV to TeV energies we see
that the emission in the SPI range cannot be easily explained as inverse Compton or
bremsstrahlung from  cosmic-ray electrons. Either another mechanism is at work
or there is a new population of compact sources with characteristic high-energy tails.

In view of the apparent difficulty to produce the very hard spectrum with a diffuse mechanism,
at present the source population  seems the most plausible explanation.
Possible candidates are the anomalous X-ray pulsars (AXPs) with the characteristic 
hardening of their spectra around 100~keV.
Detailed evaluation of the contribution from unresolved point-sources will
require population synthesis studies, and this will be the subject of future work.
Meanwhile we are investigating further the possibility of diffuse mechanisms like
that proposed by \citet{dogiel02b}.

\appendix
 %______________________________________________________________
%%%%%%%%%%%%%%%%%%%%%%%%%%%%%%%%%%%%%%%%%%%%%%%%%%%%%%%%%%%%%%%%%%%%%%%%%%%%%%%%
\section {Model Fitting Details}

The objective of the model fitting analysis is now to extract information about $\thetai$
in the form of posterior probability distributions, and their moments 
(mean, standard deviation etc) and any other functions of interest (e.g. the total image).

\subsubsection*{Fitting and error estimation}

The likelihood function is:

$$L(D|\bar\theta) = \prod_k e^{-\dk} \dk^{\nk} / \nk !$$
where $\nk$ are the measured data (denoted collectively by $D$).

and the posterior probability $P(\bar\theta|D)$ is expressed in terms of the likelihood 
and the prior probability $Pr(\bar\theta)$  using Bayes theorem:

$$P(\bar\theta|D)=L(D|\bar\theta) Pr(\bar\theta)/P(D)$$

%where P(D) is known as the {\it evidence}.

The posterior for one parameter is obtained by marginalizing over the other
parameters:
$$P(\thetai|D)=\int_{i^\prime\ne i} P(\bar\theta_i^\prime|D) d^{N}\theta^\prime$$
and its mean value is

$$<\thetai|D>=\int \thetai P(\thetai|D) d\thetai =\int \thetai P(\bar\theta|D) d^{N}\theta $$

with standard deviation

$$\Delta\thetai|D=sqrt\int (\thetai- <\thetai|D>)^2 P(\thetai|D)d\thetai$$
$$=sqrt \int   (\thetai- <\thetai|D>)^2 P(\bar\theta|D) d^{N}\theta$$

An analytical approximation to the covariance matrix and hence the 
error estimates can be obtained by expressing the
log-likelihood function as an expansion about the maximum:
$$\logL(\bar\theta)=\logL (\bar\theta_o) + {1\over2}\Sigma_p\Sigma_q (\dthetap)(\dthetaq)\Hessianpq  $$
where the Hessian matrix is $\Hessianpq = \partial^2 \logL / \partial\thetap\partial\thetaq$.
The marginalization integration can then be done analytically using diagonalization (\citet{sivia97}: Ch 3.2 and Appendix A.3)
which gives  the result for the covariance under the posterior:
$$\sigmapq^2=<(\thetap-\thetap^o)(\thetaq-\thetaq^o)>= \Hessianpq^{-1}$$
The inverse of $\Hessian$ can be written in terms of  the matrix of eigenvectors $X$ and eigenvalues $\Lambda = diag(\lambda_q)$ :
$$\Hessian X= X \Lambda$$
$$ X= \Hessian^{-1} X \Lambda$$
$$ \Hessian^{-1}=  X \Lambda^{-1} X^{-1} =  X \Lambda^{-1} X^T $$ since eigenvectors of a real symmetric matrix
are orthogonal, $  X^{-1}=X^T $. 
For a single parameter,
$$\sigmapp^2= <(\thetap-\thetap^o)^2>= \Hessianpp^{-1} =  (X \Lambda^{-1} X^T)_{pp} = \Sigma_q{ X_{pq}^2\over\lambda_q}$$
It is interesting to note that a completely different formalism involving the
distribution of the maximum-likelihood ratio \citep{strong85} and maximizing the likelihood under constraints instead of 
marginalizing over unwanted parameters,  actually leads to
the same formula for errors on one parameter,
 once we express the inverse Hessian in terms of eigenvalues.
This has not been noted before to the author's knowledge. 
One advantage of the Bayes formalism is that the covariances can be used to compute
the errors on linear combinations of components.
%NB Needs exact formulation of equivalence for 1$\sigma$ relative to deltaL).

The RMS error on a linear combination of parameters is directly related to the covariances as follows
\footnote{
This is true generally, so if we have a better approximation to $\sigmaij^2$
than the Hessian one (e.g. via Monte Carlo Markov Chain) we get a correspondingly better estimate
of $<(\Delta f)^2>$.}
$$f=\Sigma w_i\thetai$$
$$\Delta f= w_i \Delta\thetai$$
$$(\Delta f)^2 =\Sigma_i\Sigma_j w_i w_j \Delta\thetai\Delta\thetaj $$

$$<(\Delta f)^2> =\Sigma_i\Sigma_j w_i w_j <\Delta\thetai\Delta\thetaj> $$
$$=\Sigma_i\Sigma_j w_i w_j \sigmaij^2$$

%%%%%%%%%%%%%%%%%%%%%%%%%%%%%%%%%%%%%%%%%%%%%%%%%%%%%%%%%%%%%%%%%%%%%%%%%%%%%%%%%%%%%%%%%%%

\section{  Fit results}

%%%%%%%%%%%%%%%%%%%%%%%%%%%%%%%%%%%%%%%%%%%%%%%%%%%%%%%%%%%%%%%%%%%%%%%%%%%%%%%%%%%%%%%%%%%

This appendix consists of Tables B1--4, containing all the SPI results plotted in the spectra,
to enable convenient use of the present work.

%_____________________________________________________________
%
% values generated by galplot_linux
%id 2087  E=18-28 number(>sigma) :    71  65  55  46  39  29  16  6  3  2
%id 2090  E=28-38 number(>sigma) :    68  64  62  57  49  31  15  7  4  2
%id 2093  E=38-48 number(>sigma) :    60  57  53  45  39  22  14  4  2  0
%id 2096  E=48-58 number(>sigma) :    48  35  28  23  18  10  3  1  0  0
%id 2100  E=58-68 number(>sigma) :    36  25  15  11  11  4  1  0  0  0
%id 2103  E=68-78 number(>sigma) :    36  30  26  16  15  9  4  0  0  0
%id 2129  E=78-88 number(>sigma) :    37  34  27  18  11  10  3  0  0  0
%id 2132  E=88-98 number(>sigma) :    24  19  11  9  6  3  0  0  0  0
%id 2112  E=98-178 number(>sigma) :    35  22  14  8  8  5  1  0  0  0
%id 2115  E=178-258 number(>sigma) :    20  9  4  0  0  0  0  0  0  0
%id 2118  E=258-338 number(>sigma) :    19  5  3  2  1  0  0  0  0  0
%id 2121  E=338-418 number(>sigma) :    5  1  0  0  0  0  0  0  0  0
%id 2123  E=418-498 number(>sigma) :    10  2  0  0  0  0  0  0  0  0
%id 2125  E=518-1.02e+03 number(>sigma) :    9  3  0  0  0  0  0  0  0  0

%%%%%%%%%%%%%%%%%%%%%%%%%%%%%%%%%%%%%%%%%%%%%%%%%%%%%%%%%%%%%%%%%%%%%%%%%%%%%%%%%%%
% INTEGRAL/SPI spimodfit results 
\begin{table}[p]

\label{variability}      % is used to refer this table in the text
\caption{SPI diffuse emission spectrum for various source variability timescales. 
* indicates selected sources
(Sco X-1, 4U1700-377, OAO 1657-4154) 
 allowed to vary on timescale of SPI pointings.
\protect\\                     % protect necessary here
$   0.25\deg<l< 29.75\deg,    330.25\deg<l<359.75\deg $,
\protect\\ 
$  -9.75\deg<b< -0.25\deg,      0.25\deg<b<  9.75\deg$ 
    }
\centering                          % used for centering table
\begin{tabular}{l l l c c}        % left and centered columns (5 columns)
\hline\hline                 
Energy range & variability & intensity                       & error  &ID  \\    % table heading 
 (keV)       & timescale           & cm$^{-2}$ sr$^{-1}$     &        &    \\    % table heading 
             & (days)           & s$^{-1}$ MeV$^{-1}$        &        &    \\    % table heading 

\hline                   

\hline
%list 227
  18.0-  28.0  & const&5.469 &  1.711 &  2481 \\
  18.0-  28.0  & 200  &4.484 &  1.580 &  2501 \\
  18.0-  28.0  & 100  &3.572 &  1.452 &  2502 \\
  18.0-  28.0  & 50   &2.870 &  1.369 &  2503 \\
  18.0-  28.0  & *    &4.341 &  2.496 &  2505 \\
\hline
  28.0-  38.0  & const&3.725 &  0.782 &  2482 \\
  28.0-  38.0  & 200  &3.112 &  0.694 &  2511 \\
  28.0-  38.0  & 100  &2.846 &  0.656 &  2512 \\
  28.0-  38.0  & 50   &2.626 &  0.629 &  2513 \\
% 28.0-  38.0  & 20   &2.302 &  0.587 &  2514 \\
  28.0-  38.0  & *    &3.295 &  0.723 &  2515 \\
\hline
  
  38.0-  48.0  & const&1.894 &  0.643 &  2483 \\
  38.0-  48.0  & 200  &1.861 &  0.421 &  2521 \\
  38.0-  48.0  & 100  &1.783 &  0.410 &  2522 \\
  38.0-  48.0  &  50  &1.778 &  0.413 &  2523 \\
  38.0-  48.0  & *    &1.826 &  0.417 &  2525 \\
\hline
  48.0-  58.0  & const&1.859 &  0.483 &  2484 \\
  48.0-  58.0  & 200  &1.958 &  0.923 &  2531 \\
  48.0-  58.0  & 100  &1.977 &  0.930 &  2532 \\
  48.0-  58.0  &  50  &1.786 &  0.904 &  2533 \\
  48.0-  58.0  & *    &1.755 &  0.769 &  2535 \\
\hline
 58.0-  68.0  &const &1.685 &  0.469 &  2485 \\
 58.0-  68.0  & 200  &1.643 &  0.466 &  2541 \\
 58.0-  68.0  & 100  &1.602 &  0.461 &  2542 \\
 58.0-  68.0  &  50  &1.559 &  0.459 &  2543 \\

\hline                                   %inserts single line
\end{tabular}
\end{table}

%%%%%%%%%%%%%%%%%%%%%%%%%%%%%%%%%%%%%%%%%%%%%%%%%%%%%%%%%%%%%%%%%%%%%%%%%%%%%%%%%%%
% INTEGRAL/SPI spimodfit results 
\begin{table}[p]

\label{fits_A}      % is used to refer this table in the text
\caption{SPI diffuse emission spectrum for \protect\\ 
$  330\deg<l< 30\deg$, $ |b|< 10\deg$    
 }
\centering                          % used for centering table
\begin{tabular}{l l l c c}        % left and centered columns (5 columns)
\hline\hline                 
Energy range & components & intensity               & error  &ID  \\    % table heading 
 (keV)       &            & cm$^{-2}$ sr$^{-1}$     &        &    \\    % table heading 
             &            & s$^{-1}$ MeV$^{-1}$     &        &    \\    % table heading 

\hline

\hline                                   %inserts single line

\hline
%list 224
 18.0-  28.0  & 1-9  &5.469 &  1.711 &  2481 \\
  28.0-  38.0  & 1-9  &3.725 &  0.782 &  2482 \\
  38.0-  48.0  & 1-9  &1.894 &  0.643 &  2483 \\
  48.0-  58.0  & 1-9  &1.859 &  0.483 &  2484 \\
  58.0-  68.0  & 1-9  &1.685 &  0.469 &  2485 \\
  68.0-  78.0  & 1-9  &1.077 &  0.232 &  2486 \\
  78.0-  88.0  & 1-9  &0.378 &  0.110 &  2487 \\
  88.0-  98.0  & 1-9  &0.863 &  0.220 &  2488 \\
  98.0- 138.0  & 1-9  &0.444 &  0.024 &  2489 \\
 138.0- 178.0  & 1-9  &0.265 &  0.030 &  2490 \\
 178.0- 258.0  & 1-9  &0.196 &  0.020 &  2291 \\
 258.0- 338.0  & 1-9  &0.134 &  0.013 &  2248 \\
 338.0- 418.0  & 1-9  &0.106 &  0.014 &  2249 \\
 418.0- 498.0  & 1-9  &0.093 &  0.016 &  2183 \\
 508.0- 514.0  & 1-9  &0.465 &  0.064 &  2207 \\
 518.0-1018.0  & 1-9  &0.018 &  0.005 &  2186 \\

\end{tabular}
\end{table}
%%%%%%%%%%%%%%%%%%%%%%%%%%%%%%%%%%%%%%%%%%%%%%%%%%%%%%%%%%%%%%%%%%%%%%%%%%%%%%%%%%%%%%

%%%%%%%%%%%%%%%%%%%%%%%%%%%%%%%%%%%%%%%%%%%%%%%%%%%%%%%%%%%%%%%%%%%%%%%%%%%%%%%%%%%
% INTEGRAL/SPI spimodfit results 
\begin{table}[p]

\label{fits_sources}      % is used to refer this table in the text
\caption{SPI total source emission spectrum for \protect\\ 
$  330\deg<l< 30\deg$, $ |b|< 10\deg$    
}
\centering                          % used for centering table
\begin{tabular}{l l l c c}        % left and centered columns (5 columns)
\hline\hline                 
Energy range & components & intensity               & error  &ID  \\    % table heading 
 (keV)       &            & cm$^{-2}$ sr$^{-1}$     &        &    \\    % table heading 
             &            & s$^{-1}$ MeV$^{-1}$     &        &    \\    % table heading 

\hline                                   %inserts single line

\hline
%list 224
  18.0-  28.0  & 10-132  &64.054 &  0.630 &  2481 \\
  28.0-  38.0  & 10-132  &18.619 &  0.161 &  2482 \\
  38.0-  48.0  & 10-106  &9.585 &  0.095 &  2483 \\
  48.0-  58.0  & 10-71  &5.004 &  0.116 &  2484 \\
  58.0-  68.0  & 10-53  &2.707 &  0.108 &  2485 \\
  68.0-  78.0  & 10-45  &1.579 &  0.036 &  2486 \\
  78.0-  88.0  & 10-50  &1.198 &  0.029 &  2487 \\
  88.0-  98.0  & 10-39  &0.770 &  0.047 &  2488 \\
  98.0- 138.0  & 10-32  &0.379 &  0.012 &  2489 \\
 138.0- 178.0  & 10-21  &0.114 &  0.010 &  2490 \\
% 178.0- 258.0  & 10-13  &0.000 &  0.000 &  2291 \\
% 258.0- 338.0  & 10-13  &0.000 &  0.000 &  2248 \\
% 338.0- 418.0  & 10-12  &0.000 &  0.000 &  2249 \\
% 418.0- 498.0  & 10-11  &0.000 &  0.000 &  2183 \\
% 518.0-1018.0  & 10-11  &0.000 &  0.000 &  2186 \\

\end{tabular}
\end{table}
%%%%%%%%%%%%%%%%%%%%%%%%%%%%%%%%%%%%%%%%%%%%%%%%%%%%%%%%%%%%%%%%%%%%%%%%%%%%%%%%%%%%%%

% INTEGRAL/SPI spimodfit results for list 212 
\begin{table}[p]
\caption{SPI diffuse emission spectrum for \protect\\ 
$  0.25\deg<l<  9.75\deg$,    $350.25\deg<l<359.75\deg$,
$ -9.75\deg<b< -0.25\deg$,    $  0.25\deg<b<  9.75\deg$
 .   }
\label{fits_B}      % is used to refer this table in the text
\centering                          % used for centering table
\begin{tabular}{l l l c c}        % left and centered columns (5 columns)
\hline\hline                 % inserts double horizontal lines
Energy range & components & intensity               & error  &ID  \\    % table heading 
 (keV)       &            & cm$^{-2}$ sr$^{-1}$     &        &    \\    % table heading 
             &            & s$^{-1}$ MeV$^{-1}$     &        &    \\    % table heading 

\hline                        % inserts single horizontal line

\hline                                   %inserts single line

\hline
%list 224
  18.0-  28.0  & 1-9  &13.829 &  3.434 &  2481 \\
  28.0-  38.0  & 1-9  &7.015 &  1.402 &  2482 \\
  38.0-  48.0  & 1-9  &4.502 &  1.065 &  2483 \\
  48.0-  58.0  & 1-9  &3.754 &  0.858 &  2484 \\
  58.0-  68.0  & 1-9  &2.335 &  0.680 &  2485 \\
  68.0-  78.0  & 1-9  &2.217 &  0.442 &  2486 \\
  78.0-  88.0  & 1-9  &0.574 &  0.173 &  2487 \\
  88.0-  98.0  & 1-9  &1.599 &  0.376 &  2488 \\
  98.0- 138.0  & 1-9  &0.825 &  0.037 &  2489 \\
 138.0- 178.0  & 1-9  &0.492 &  0.043 &  2490 \\
 178.0- 258.0  & 1-9  &0.359 &  0.028 &  2291 \\
 258.0- 338.0  & 1-9  &0.237 &  0.019 &  2248 \\
 338.0- 418.0  & 1-9  &0.190 &  0.019 &  2249 \\
 418.0- 498.0  & 1-9  &0.190 &  0.022 &  2183 \\
 508.0- 514.0  & 1-9  &1.244 &  0.095 &  2207 \\
 518.0-1018.0  & 1-9  &0.022 &  0.008 &  2186 \\

\end{tabular}
\end{table}
%%%%%%%%%%%%%%%%%%%%%%%%%%%%%%%%%%%%%%%%%%%%%%%%%%%%%%%%%%%%%%%%%%%%%%%%%%%%%%%%%%%%%%
\begin{table}[p]
%\caption{Spectra of diffuse emission from model fitting.}             % title of Table

\caption{SPI diffuse emission spectrum for 
\protect\\ 
$  9.25\deg < l <  29.75\deg$,    $330.25\deg < l < 350.75\deg$,
$ -9.75\deg < b <  -0.25\deg$,    $  0.25\deg < b <   9.75\deg$
   }
\label{fits_C}      % is used to refer this table in the text
\centering                          % used for centering table
\begin{tabular}{l l l c c}        % left and centered columns (5 columns)
\hline\hline                 % inserts double horizontal lines
Energy range & components & intensity               & error  &ID  \\    % table heading 
 (keV)       &            & cm$^{-2}$ sr$^{-1}$     &        &    \\    % table heading 
             &            & s$^{-1}$ MeV$^{-1}$     &        &    \\    % table heading 

\hline                        % inserts single horizontal line

\hline                                   %inserts single line

\hline
%list 224
  18.0-  28.0  & 1-9  &1.331 &  1.179 &  2481 \\
  28.0-  38.0  & 1-9  &2.147 &  0.564 &  2482 \\
  38.0-  48.0  & 1-9  &0.610 &  0.455 &  2483 \\
  48.0-  58.0  & 1-9  &0.936 &  0.369 &  2484 \\
  58.0-  68.0  & 1-9  &1.363 &  0.441 &  2485 \\
  68.0-  78.0  & 1-9  &0.503 &  0.154 &  2486 \\
  78.0-  88.0  & 1-9  &0.286 &  0.101 &  2487 \\
  88.0-  98.0  & 1-9  &0.494 &  0.173 &  2488 \\
  98.0- 138.0  & 1-9  &0.252 &  0.027 &  2489 \\
 138.0- 178.0  & 1-9  &0.152 &  0.033 &  2490 \\
 178.0- 258.0  & 1-9  &0.119 &  0.022 &  2291 \\
 258.0- 338.0  & 1-9  &0.083 &  0.015 &  2248 \\
 338.0- 418.0  & 1-9  &0.064 &  0.015 &  2249 \\
 418.0- 498.0  & 1-9  &0.046 &  0.017 &  2183 \\
 508.0- 514.0  & 1-9  &0.085 &  0.072 &  2207 \\
 518.0-1018.0  & 1-9  &0.016 &  0.006 &  2186 \\

\end{tabular}
\end{table}
\begin{acknowledgements}
We thank the IBIS Survey Team for making available a preliminary version of the 2nd ISGRI catalogue prior to publication.
We thank Michael Revnivtsev for providing his RXTE spectrum and help with the normalization.

This paper is based on observations with INTEGRAL, an ESA project with instruments and a science data center funded by ESA member states (especially the PI countries: Denmark, France, Germany, Italy, Switzerland, Spain), Czech Republic and Poland, and with the participation of Russia and the USA. The SPI project has been completed under the responsibility and leadership of CNES/France. The SPI anticoincidence system is supported by the German government through DLR grant 50.0G.9503.0. We are grateful to ASI, CEA, CNES, DLR, ESA, INTA, NASA and OSTC for support.

 \end{acknowledgements}
%%%%%%%%%%%%%%%%%%%%%%%%%%%%%%%%%%%

\end{document}